\documentclass[12pt]{iopart}
\makeatletter
\expandafter\let\csname equation*\endcsname\relax
\expandafter\let\csname endequation*\endcsname\relax
\usepackage[utf8]{inputenc}
\usepackage{color}
\usepackage{enumerate}
\usepackage{graphicx}
\usepackage{bm}
\usepackage{qcircuit}
\usepackage{hyperref}
\usepackage{physics}
\usepackage{subfigure}
\usepackage{enumerate}
\usepackage{adjustbox}
\usepackage{indentfirst}
\usepackage{graphicx}
\usepackage{float}

\usepackage{listings}
\usepackage{xcolor}
\hypersetup{  colorlinks=true, linkcolor=blue, citecolor=blue, urlcolor=blue  }
\newcommand{\red}{\textcolor{red}}
\newcommand{\blue}{\textcolor{blue}}
\newcommand{\mol}[1]{\mathrm{#1}}
\DeclareUnicodeCharacter{2212}{-}

\begin{document}

\title{Benchmarking Variational Quantum Eigensolvers for Quantum Chemistry}

\author{MindSpore Quantum Developers\footnote{\textbf{Corresponding author. Email:} copperhu@foxmail.com (J.Hu); yung@sustech.edu.cn (M.-H.Yung)}\footnote{Full author list and author contributions in the acknowledgement section.}}
 \address{Central Research Institute, 2012 Labs, Huawei Technologies}
 \address{Department of Physics, Southern University of Science and Technology, Shenzhen 518055, China}
 \address{Shenzhen Institute for Quantum Science and Engineering, Southern University of Science and Technology, Shenzhen 518055, China}
 \address{LINKE Lab, School of Computer Science and Technology, University of Science and Technology of China, Hefei 230027, China}
 \address{Key Laboratory of Wireless-Optical Communications, University of Science and Technology of China, Chinese Academy of Sciences, Hefei 230027, China}


\begin{abstract}
  Quantum chemistry is one of the most promising applications of quantum computers in the near future. For noisy intermediate-scale quantum devices, the quantum-classical hybrid framework based on the variational quantum eigensolver (VQE) has become the method of choice. In the literature, there are many different variants of VQE, but it is not known which one is optimal for a given molecule. For this purpose, we perform a thorough benchmarking on more than ten different kinds of VQE ansatzes (in systems up to 30 qubits), based on their performance on the energy accuracy, runtime until convergence, and number of parameters. Our results show that the ADAPT ansatz can be used to obtain more accurate energy for small systems (below 14 qubits), but it costs much more computational resources. For larger molecules, UCCSD0 has better performance. However, all the tested ansatzes can hardly reach chemical accuracy at stretched bond lengths. Our results were obtained using MindSpore Quantum, where the codes and the benchmarking toolkit are publicly available at Gitee.
  
\end{abstract}

\tableofcontents

\maketitle

\section{Introduction}
  Solving the ground-state problem of many-electrons molecular systems is of immense importance for physics and chemistry. This problem is computational challenging and usually needs to be solved by approximate methods.
  When the electron-electron interaction is weak, one can apply the Hartree-Fock (HF) method, which is essentially a mean-field approximation. When the electron-electron interaction is strong, 
  one may employ some classical methods such as coupled cluster~(CC)\cite{HOFMANN2003487}, and configuration interaction~(CI)\cite{BELLAYOUNI2014203}. Moreover, without any truncation of the CI method, the full configuration interaction~(FCI) approach can give exact ground energy of the system of interest on a specific basis. However, to accurately simulate the molecules, the computational cost of these methods scales exponentially with the size of the system\cite{THOGERSEN200436}. Such exponential issue hinders their applications to larger molecular systems.
  
  Quantum computing is supposed to be a promising way of overcoming the above issue, as proposed by Feynman in 1982~\cite{feynman1982}. Since then, various quantum algorithms have been developed for simulations~\cite{aspuru2005simulated, kassal2008polynomial, Huh2014a, peruzzo2014variational}. Among them, the variational quantum eigensolver (VQE)~\cite{peruzzo2014variational, yung2014transistor} is believed to be friendly to near-term quantum devices for its noise-resilient property and shallower quantum circuits~\cite{o2016scalable}.
  VQE is a hybrid quantum-classical variational algorithm, including two parts that will be implemented in quantum and classical computers, respectively. The quantum computer prepares the required quantum states with designed quantum circuits, in which include quantum gates with parameters to be optimized. Then corresponding measurement is made with the states. Measurement results are gathered and sent to the classical computer for parameter optimization, then the optimized parameters will be sent back to the quantum computer, to start the next iteration.
  
  VQE shows an optimistic prospect in recent researches. Since the first $\mol{HHe^+}$ (2 qubits) demonstration presented in Ref.~\cite{peruzzo2014variational}, a variety of experimental researches with VQE are proposed in small molecular systems,
  like $\mol{BeH_2}$ (6 qubits)~\cite{Kandala2017}, $\mol{H_2O}$ (8 qubits)~\cite{Nam2020} and $\mol{H_{12}}$ (12 qubits)~\cite{AIQuantum2020}. A variety of works have focused on designing different types of ansatzes~\cite{Kandala2017, Ryabinkin2020, Grimsley, Dallaire_Demers_2019, wecker2015progress}, showing that different variants of the VQE method may be capable in different systems. However, which ansatz is more suitable for different molecular systems remains unknown.
  
  There have been some researches benchmarking the performance of different VQE settings in quantum simulators~\cite{Yeter-Aydeniz, Lolur2021, Kuhn2019}, but the existing results only included very few ansatzes. Besides, a toolkit that can make comparisons between various ansatzes and record all the results is needed.
  
  In order to make it clear which settings of VQE are more suitable for different molecular systems, in this project, we reproduced several popular ansatzes and evaluate their performance. We have benchmarked various VQE ansatzes for $\mol{H_4}$ (8 quibts), $\mol{LiH}$ (12 qubits), $\mol{BeH_2}$ (14 qubits), $\mol{H_2O}$ (14 qubits), $\mol{CH_4}$ (18 qubits), and $\mol{N_2}$(20 qubits) in STO-3G basis set and compared their energy error, convergence runtime, and number of parameters in the circuits. Furthermore, we analyzed the results using the toolkit we developed, and drew brief conclusions about their advantages and disadvantages when simulating different systems. These results be helpful for researchers to use VQE in practice for different molecular systems. The codes and results are available in \cite{MindQuantum}.
  
  The article is structured as follows: Section~\ref{sec:VQE} will introduce the VQE method; Section~\ref{sec:ansatz} will briefly introduce different variational ansatzes we have benchmarked; Section~\ref{sec:tools} will introduce the benchmarking tools used in the project. Section~\ref{sec:results} will show the benchmarking results, including the comparison of energy error, convergence runtime, and number of parameters in the circuits between different ansatzes. Section~\ref{sec:discussion} will discuss how will the project develop in the future.

\section{Variational Quantum Eigensolver Framework}\label{sec:VQE}
    
    In this section, we will overview the scheme of the VQE method (Sec~\ref{ssec:veq_proc}), as well as introduce some key technical details of implementation. The implementation details include the molecular Hamiltonian buildup (Sec~\ref{ssec:mol}), the fermion-to-qubit mapping (Sec~\ref{ssec:mapping}), the Trotterization for the time evolution operator (Sec~\ref{ssec:trotter}) and the parameter optimization (Sec~\ref{ssec:param_optim}).

    \subsection{VQE Scheme}\label{ssec:veq_proc}
    VQE algorithm is designed to obtain the ground state of quantum systems. It is based on the Rayleigh-Ritz variational principle as 
    \begin{equation}
        \langle\Psi(\boldsymbol{\theta})|\hat{H}|\Psi(\boldsymbol{\theta})\rangle=\langle E\rangle \geq {E_{g}},
        \label{rr}
    \end{equation}
    where $\langle E\rangle$ is the expectation value of the Hamiltonian $\hat{H}$; $\ket{\Psi(\boldsymbol{\theta})}$ is the approximated target state prepared by applying the quantum circuit $\hat{U}(\boldsymbol{\theta})$ on the initial state $\ket{\Psi_0}$; $E_{g}$ is the exact ground state energy of the system. 
    
\begin{figure}[htbp]
        \centering
        \includegraphics[width=\linewidth]{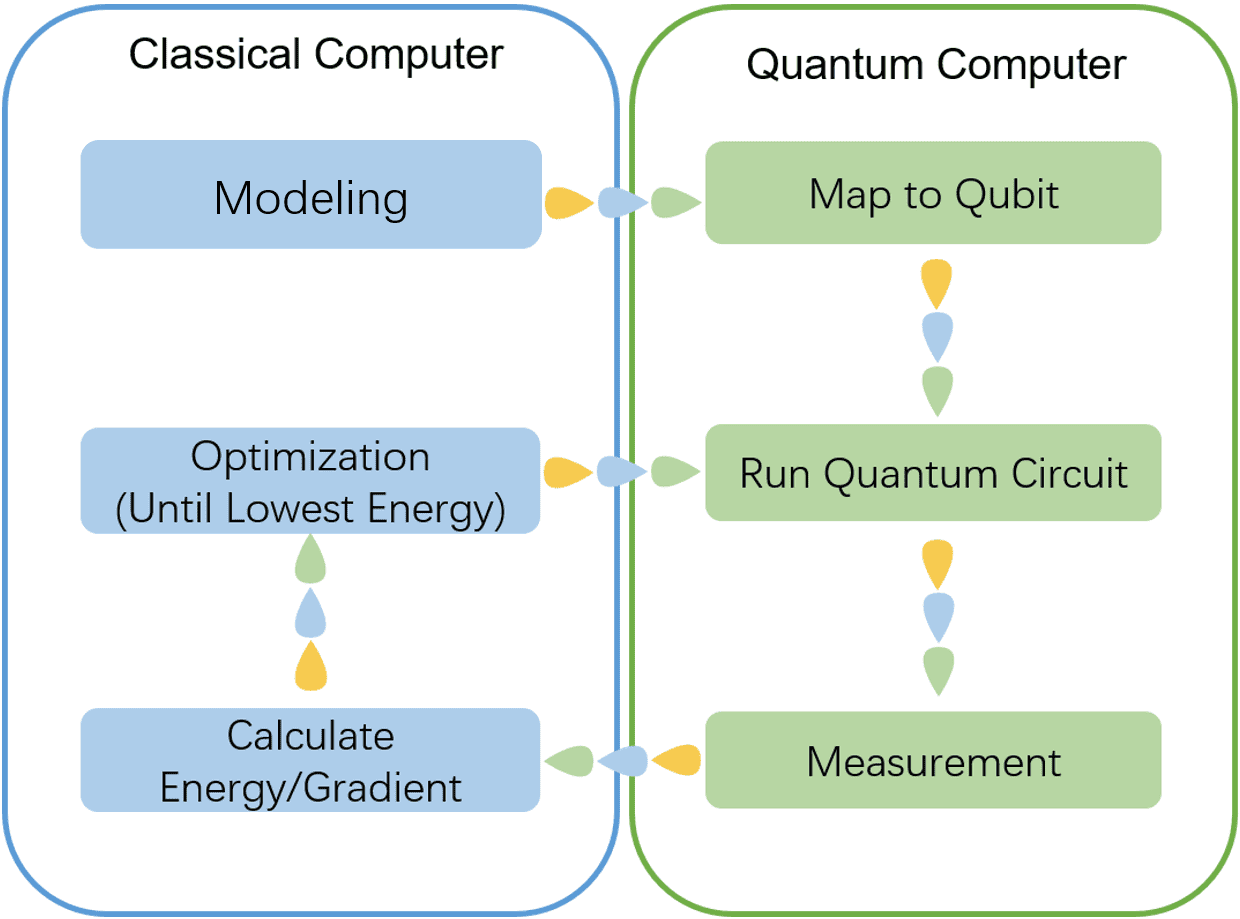}
        \caption{The scheme of VQE. Molecule modeling, parameter optimization, and energy calculation steps are executed on a classical computer (left side). Quantum computers are responsible for the state evolution part (right side). The bottom loop will be terminated when the condition of convergence is satisfied.}
        \label{VQE-Scheme}
    \end{figure}

    A general VQE scheme for molecular simulation contains the core steps below 
    and is also illustrated in Figure~\ref{VQE-Scheme}.
    \begin{enumerate}[(1)]
        \item Modeling: Input the geometry of moleucles, including three dimensional coordinates of each atom. Compute one- and two-electron integrals of molecules of a predefined basis set and a second-quantized Hamiltonian can be obtained. (See subsection~\ref{ssec:mol})
        
        \item Map to qubit: Transform the second-quantized
        Hamiltonian to a qubit form by a fermion-to-qubit encoding.(See subsection~\ref{ssec:mapping})
        
        \item Run Quantum circuit: also state preparation step. A trial wave function $|\Psi(\boldsymbol{\theta})\rangle$, called ansatz, is prepared by a parameterized quantum circuit (PQC) $\hat{U}(\boldsymbol{\theta})$ and an initial state $\ket{\Psi_0}$.
        Suitable ansatzes is necessary for the study of models and the capability of quantum hardwares.
        \item Measurement:
        State preparation and measurement are repeated many times to evaluation the expectation value of observables. Typically a product of Pauli operators. (See subsection \ref{ssec:mol})
        \item Calculate energy/gradient: The expectation value $\langle E \rangle$ of the Hamiltonian $\hat{H}$ is calculated by the Hamiltonian averaging step, which includes measurements of each tensor product of Pauli operators to get expectation values and adding them together. 
        Gradients of parameters are necessary for many optimization methods. Calculation of gradients is employed by effective approaches including finite difference, analytical gradients. However,
        gradients of parameters in this paper can be conveniently returned by MindQuantum\cite{MindQuantum} with the ability of automatic differentiation.
        
        \item Optimization: This step is carried out by classical computers. Parameters in the ansatz $\ket{\Psi(\boldsymbol{\theta})}$ are updated by optimizing methods to minimize the expectation value of the problem Hamiltonian. (See subsection \ref{ssec:param_optim})
    \end{enumerate}
    
    \subsection{Molecular Hamiltonian}\label{ssec:mol}
    The Hamiltonian of a molecule with $N_e$ electrons and $N_n$ nuclei can be written as:
    \begin{equation}
        \begin{aligned}
            \hat{H} &= \sum_{i}^{N_e} -\frac{\nabla_i}{2} + \sum_{i<j}^{N_e}\frac{1}{|\mathbf{r}_i-\mathbf{r}_j|} +  \sum_{I<J}^{N_n}\frac{Z_I Z_J}{|\mathbf{R}_I-\mathbf{R}_J|} \\
            &+ \sum_{i,I}\frac{Z_I}{|\mathbf{r}_i-\mathbf{R}_I|},
        \end{aligned}
        \label{r_h}
    \end{equation}
    where $\mathbf{R}, \mathbf{r}$ is the three-dimension coordinates of nuclei and electrons, respectively. $Z$ represents the charges of nuclei. The Born-Oppenheimer (BO) approximation is implicitly specified in Equation~\ref{r_h}, which neglects the nuclear motion and treats them as fixed points.
     Typically, the problem Hamiltonian can be written in a second-quantized form in the predefined basis set as 

    \begin{equation}
        \hat{H}=\sum\limits_{p,q} h_{pq}\hat{a}_p^{\dagger}\hat{a}_q
    	+\frac{1}{2}\sum\limits_{p,q,r,s} g_{pqrs}\hat{a}_p^{\dagger}\hat{a}_r^{\dagger}\hat{a}_{s} \hat{a}_{q},
    	\label{sq_ham}
    \end{equation}
    where $\hat{a_p}^\dag$ and $\hat{a_p}$ denote the fermionic creation operator and annihilation operator associated with $p$-th fermionic mode (or spin orbital). The sets of coefficients $\{h_{pq}\}$ and $\{g_{pqrs}\}$ are one- and two-electron integrals which can be evaluated by classical computers. 
    
    The second-quantized Hamiltonian in Equation~\ref{sq_ham} has to be transformed by a fermion-to-qubit mapping (encoding) to enable calculations on quantum computers. The Hamiltonian after mapping is a linear summation of Pauli strings as
    \begin{equation}
        \hat{H} = \sum_{i} c_i\hat{P}_i=\sum_{i} c_i \prod_i \sigma_j^i,
        \label{q_ham}
    \end{equation}
    where $c_i$ is the real-valued coefficient, $\hat{P}_i$ denotes $i$-th Pauli string of the Hamiltonian, which contains a series of Pauli operators, and $\sigma_j^i$ is one of such pauli operators $\{\hat{\sigma}^I, \hat{\sigma}^x, \hat{\sigma}^y, \hat{\sigma}^z\}$ acting on $j$-th qubit. 
    
    By the Hamiltonian averaging step, the expectation value of the Hamiltonian can be evaluated as
    \begin{equation}
        E(\boldsymbol{\theta}) = \sum_{i} c_i \langle\Psi(\boldsymbol{\theta})|\prod_i \sigma_j^i|\Psi(\boldsymbol{\theta})\rangle.
        \label{measure-outcome}
    \end{equation}
    In this step, the quantum state has to be prepared and measured many times to reach the required precision. With the linearity of the Hamiltonian, the expectation value of each term can be performed in parallel.
    
    \subsection{Mapping}\label{ssec:mapping}
    To perform quantum chemistry simulation in the second-quantized form on a quantum computer, it is necessary to transform fermionic operators to operators that act on qubits. The mapping enables states in the Fock space to be represented by qubit states in the Hilbert space. There are many different mappings such as Jordan-Wigner\cite{Jordan1928} and Bravyi-Kitaev\cite{BRAVYI2002210}. In this study, we will focus on the Jordan-Wigner mapping. 
    
    The Jordan-Wigner mapping stores the occupation information of each spin orbital with qubit state $\ket{0}$ and $\ket{1}$ as
     \begin{equation}
         \begin{aligned}
             \ket{f_0, f_1, \cdots, f_i, \cdots, f_n} \longrightarrow \ket{q_0} \otimes \cdots \otimes \ket{q_n},
             \label{jwencoding}
         \end{aligned}
     \end{equation}
    where the $n$-mode occupation number vector in the Fock space is written as $\ket{f_0, f_1, \cdots, f_i, \cdots, f_n}$, where $f_i \in \{0, 1\}$ is the occupation number and $0 (1)$ denotes the $i$-th spin orbital is unoccupied (occupied).
     
    Next, the fermionic creation and annihilation operators are transformed to qubit operators. For $\hat{a}^\dagger_i$ and $\hat{a}_i$ acting on $i$-th spin orbital, we have
    \begin{equation}
        \begin{aligned}
            \hat{a}^\dagger_i \longrightarrow \hat{Q}^\dagger_i \otimes \hat{\sigma}^z_{i-1} \otimes \hat{\sigma}^z_{i-2}  \cdots \otimes \hat{\sigma}^z_{0}, \\
            \hat{a}_i \longrightarrow \hat{Q}_i \otimes \hat{\sigma}^z_{i-1} \otimes \hat{\sigma}^z_{i-2}  \cdots \otimes \hat{\sigma}^z_{0},
            \label{jw}
        \end{aligned}
    \end{equation}
    where $\hat{Q}^\dagger = \ket{1}\bra{0} = (\hat{\sigma}^x-i\hat{\sigma}^y)/2$ and $\hat{Q} = \ket{0}\bra{1} = (\hat{\sigma}^x+i\hat{\sigma}^y)/2$. Jordan-wigner mapping represent the action of $\hat{a}^\dagger_i$ ($\hat{a}_i$) with $\hat{Q}^\dagger_i$ ($\hat{Q}_i$) and a series of Pauli Z operators acting on qubits of which index less than $i$. The sequence of Pauli Z operators will give a extra phase factor $-1$ if the parity of qubits with index less than $i$ is odd. This step preserve the anti-commutation relation of fermionic operators. It is worth mentioning that a single fermionic operator requires $O(n)$ qubit operations with Jordan-Wigner mapping in Equation~\ref{jw}.
    
    \subsection{Trotterization}\label{ssec:trotter}
    
    Trotter–Suzuki decompositions\cite{SUZUKI1990319} are widely used to approximate the time evolution operator $e^{i\hat{H}t}$. With the Trotter–Suzuki formula, $e^{i\hat{H}t}$ is divided into products of individual terms and then can be conveniently transformed to circuits in quantum computers. The trotter error can be compressed by increasing the trotter number $N$.

    The first-order Trotter–Suzuki approximation is written as:
    \begin{equation}
        e^{i\hat{H}t} \approx (\prod_{k=1}^{n}e^{i\hat{H}_k t/N})^N,
    \end{equation}
    Where $\hat{H}=\sum_{k=1}^n \hat{H}_k$. Practically, the trotter number N is usually set as 1 to reduce qubit operations. More accurate approximation can be obtained by the second-order Trotter–Suzuki formula:
    \begin{equation}
        e^{i\hat{H}t} \approx (\prod_{k=1}^{n}e^{i\hat{H}_k t/2N} \prod_{k=n}^{1}e^{i\hat{H}_k t/2N})^N.
    \end{equation}
    The second-order Trotter–Suzuki formula needs more operations to make a better approximation. In this paper, the first-order Trotter–Suzuki approximation is employed and the trotter number $N$ is set as $1$.
    
    \subsection{Parameter optimization}\label{ssec:param_optim}
    Parameters optimizing is of great importance for the VQE scheme. There are two main problems in this part. The first problem is how to initialize all parameters properly. Typically we can set all parameters as constant values such as $0$ or randomly sample parameters in $[0, 2\pi)$. The second problem is the choice of classical algorithms, which can be divided into gradient-based and gradient-free methods. Here we choose a gradient-based algorithm named BFGS (Broyden–Fletcher–Goldfarb–Shannon) in Scipy\cite{2020SciPy-NMeth}. More algorithms will be explored in the future.
    
    Many strategies have been proposed to address the issue of optimizing parameters in ansatzes. For example, it is a poor choice to initialize the whole parameters randomly in Hardware efficient ansatz because of the Barren Plateau phenomenon\cite{McClean2018}. To solve the problem, methods such as layer-wise learning\cite{Skolik2021} and identity block strategy\cite{Grant2019} are proposed. For chemically inspired UCCSD ansatz, a guess from second-order Møller–Plesset perturbation theory (MP2) amplitudes\cite{Romero2019} is a helpful choice.
    
\section{Variational Ansatzes} \label{sec:ansatz}
  
  Researchers have developed many ansatzes in VQE for solving molecular Hamiltonians. We picked up several representative ansatzes, shown in Table~\ref{tab:ansatzes}, and further evaluated them in this paper.

\begin{table*}[tbp] 
\footnotesize
\centering
\setlength{\tabcolsep}{2.5mm}{
\begin{tabular}[H]{llll} 
\hline
Ansatz & Particle Conservation & Circuit Type & Comments \\
\hline
UCCSD\cite{peruzzo2014variational} & Yes & Fixed & Unitary coupled cluster single and double \\ & & & ansatz. Inspired by classical CCSD methods. \\ & & &\\
UCCSD0\cite{Sokolov} & Yes & Fixed & Another parameterization version \\ & & & of UCCSD ansatz. \\ & & &\\
HEA\cite{Kandala2017} & No & Layered & Hardware efficient ansatz. Using omnipotent \\ & & &  rotation gate group and linear entanglement. \\ & & &\\
k-UpCCGSD\cite{Lee2019} & Yes & Fixed & Similar to UCCSD, but only including \\&&& paired  double excitations. One Layer \\&&& of the circuit repeats k times.\\ & & &\\
QUCC\cite{Yordanov2020} & Yes & Fixed & Qubit unitary coupled cluster ansatz. Using\\ & & & qubit excitation operators rather than orbital \\&&&
excitation operators. \\ & & &\\
ADAPT\cite{Grimsley} & Yes & Adaptive & Adaptive, derivative-assembled, pseudo-\\ & & & Trotter VQE ansatz. Adaptively add evolution \\ & & & of excitation  operators to construct ansatz. \\& & &\\
QCC\cite{Ryabinkin2018, Ryabinkin2020} & No & Adaptive & Qubit coupled cluster ansatz. Adaptively add \\ &&& pauli evolution operators to construct ansatz. \\& & &\\
qubit-ADAPT\cite{Tang2019} & No & Adaptive & Similar to QCC, but the operators are  \\ & & & only Pauli operators transformed from \\ & & & the Hamiltonian. \\ & & &\\

LDCA\cite{Dallaire_Demers_2019} & No & Layered & Low depth circuit ansatz. Motivated by \\ & & & the Bogoliubov coupled cluster theory, \\ & & & using the theory of matchgates. \\ & & &\\

BRC\cite{AIQuantum2020} & Yes & Layered & Basis rotation circuit ansatz. Realizing  \\ & & & noninteracting fermionic evolutions \\ & & & by repeating Givens rotation. \\ & & &\\
\hline
\end{tabular}
}
\caption{A summary of all ansatzes reviewed in this article. The particle conservation column indicates whether the total number of particles is conserved with the corresponding ansatz.
The circuit type column shows feature of the ansatz structure, which is fixed, layered or adaptive.} \label{tab:ansatzes}
\end{table*}

  \subsection{Unitary Coupled Cluster Singles and Doubles Ansatz (UCCSD)}
    The Unitary Coupled Cluster (UCC) ansatz has been widely used from the very beginning of the VQE study. It is inspired by classical ansatz\cite{Taube2006}. UCC ansatz can be written as
    \begin{equation}
        |\Psi(\boldsymbol{\theta})\rangle= e^{\hat{T}(\boldsymbol{\theta})-\hat{T}^\dagger(\boldsymbol{\theta})}|\Psi_0\rangle ,
    \end{equation}
    where $\left|\Psi_{0}\right\rangle$ is an initial state, usually the Hartree-Fock state. $\hat{T}(\boldsymbol{\theta})$ is the coupled-cluster excitation operator
    and can be written as
    \begin{equation}\label{eq:ucc}
        \hat{T}(\boldsymbol{\theta})=\sum\limits_{k}\hat{T}_{k}(\boldsymbol{\theta}) ,
    \end{equation}
    where k refers to the excitation level and can be truncated to single and double excitations in practice:


    \begin{equation}
        \hat{T}_{1}(\boldsymbol{\theta})=
        \sum\limits_{i,a}\hat{t}_i^a=
        \sum\limits_{i,a}\theta_i^a\hat{a}_a^{\dagger}\hat{a}_i,
    \end{equation}

    \begin{equation}
        \hat{T}_{2}(\boldsymbol{\theta})=\sum\limits_{i,j, a, b}\hat{t}_{ij}^{ab}=
        \sum\limits_{i,j,a,b}\theta_{ij}^{ab}\hat{a}_a^{\dagger}\hat{a}_b^{\dagger}\hat{a}_i \hat{a}_j,
    \end{equation}
    where $\hat{a}$ and $\hat{a}^\dagger$ are annihilation and creation operators. As a default index of UCCSD, index $i$ and $j$ denote occupied spin orbitals, $a$ and $b$ denote unoccupied spin orbitals. 
    $\theta$ is a group of parameters, which will be optimized during the variational algorithm.
    Note that the form $\hat{U}(\boldsymbol{\theta})=e^{\hat{T}(\boldsymbol{\theta})-\hat{T}^\dagger(\boldsymbol{\theta})}$ ensures the operator to be unitary. 

    In this project, we truncated the operators in Equation.~\ref{eq:ucc} to $k = 2$ and only include the excitations that preserve the singlet spin. This method is called singlet-UCCSD in our codes\cite{MindQuantum}. In singlet-UCCSD, a double excitations can be seen as two single excitations, and the spatial orbitals that be involved of these two single excitations will effect the parameter of the double excitation.
    With such parameterization method, the number of parameters can be reduced without resulting much energy error.
    
    \subsection{Singlet-paired Unitary Coupled Cluster Singles and Doubles (UCCSD0)}
      Singlet-paired UCC (UCCSD0) is an ansatz inspired by the parameterization method CCD0 ansatz\cite{Bulik2015} in classical computational chemistry. The UCCSD0 ansatz also conserves spin as that in the singlet-UCCSD ansatz. And the excitation operators are the same as that in the UCCSD ansatz. However, some parameters ahead of the double excitation operators in UCCSD0 ansatz may be different. In UCCSD0, the double excitations will be seen as a pair of electrons being excited to higher orbitals, which can be represented as
      \begin{equation}
          \hat{T}_{2}=T_{2}^{[0]}+T_{2}^{[1]} 
      \end{equation}
      and
      \begin{equation}
       \begin{aligned}
        &T_{2}^{[0]}=\frac{1}{2} \sum_{i j a b} \sigma_{i j}^{a b} P_{a b}^{\dagger}     P_{i j}, \\
        &T_{2}^{[1]}=\frac{1}{2} \sum_{i j a b} \pi_{i j}^{a b} \vec{Q}_{a b}^{\dagger}     \vec{Q}_{i j}, 
      \end{aligned}
     \end{equation}
      where $i,j$ denote the occupied spatial orbitals, $a,b$ denote the unoccupied spatial orbitals; $P, Q$ denote two types of pair annihilation operators. And the parameters $\sigma_{i j}^{a b}, \pi_{i j}^{a b}$ in UCCSD0 can be connected to the UCCSD ansatz by a specific way, which can be referred to the article\cite{Sokolov} and the corementation can be found in MindQuantum\cite{MindQuantum}.
        
    \subsection{Hardware Efficient Ansatz (HEA)}
      Hardware Efficient Ansatz (HEA) is designed for recent quantum devices\cite{Kandala2017}. The corresponding quantum circuit of HEA contain D repeated layers. Each layer includes entangling gates and single-qubit rotations with different parameters, respectively. The ansatz can be formulated as
      \begin{equation}\label{eq:hea_ansatz}
      |\Psi(\bm{\theta})\rangle=U^{D}\left(\bm{\theta^{D}}\right) U_{e n t} \cdots U^{1}\left(\bm{\theta^{1}}\right) U_{e n t} U^{0}\left(\bm{\theta^{0}}\right)|\Psi_0\rangle ,
      \end{equation}
      where $U^{L}\left(\bm{\theta}^{L}\right) =U^{L}_n\left(\bm{\theta}^{L}_n\right) \cdots U^{L}_1\left(\bm{\theta}^{L}_1\right) U^{L}_0\left(\bm{\theta}^{L}_0\right)$, and $U^{L}_i\left(\bm{\theta}^{L}_i\right)$  denotes the single-qubit rotations implemented for the i-$th$ qubit in the L-$th$ layer. $U_{ent}$ denotes entangling gates without parameters sandwiched between two single-qubit rotations. The total circuit contains D layers, where each layer contains these two parts. 
      
      \begin{figure}[htbp]
          \centering
          \includegraphics[width=0.9\linewidth]{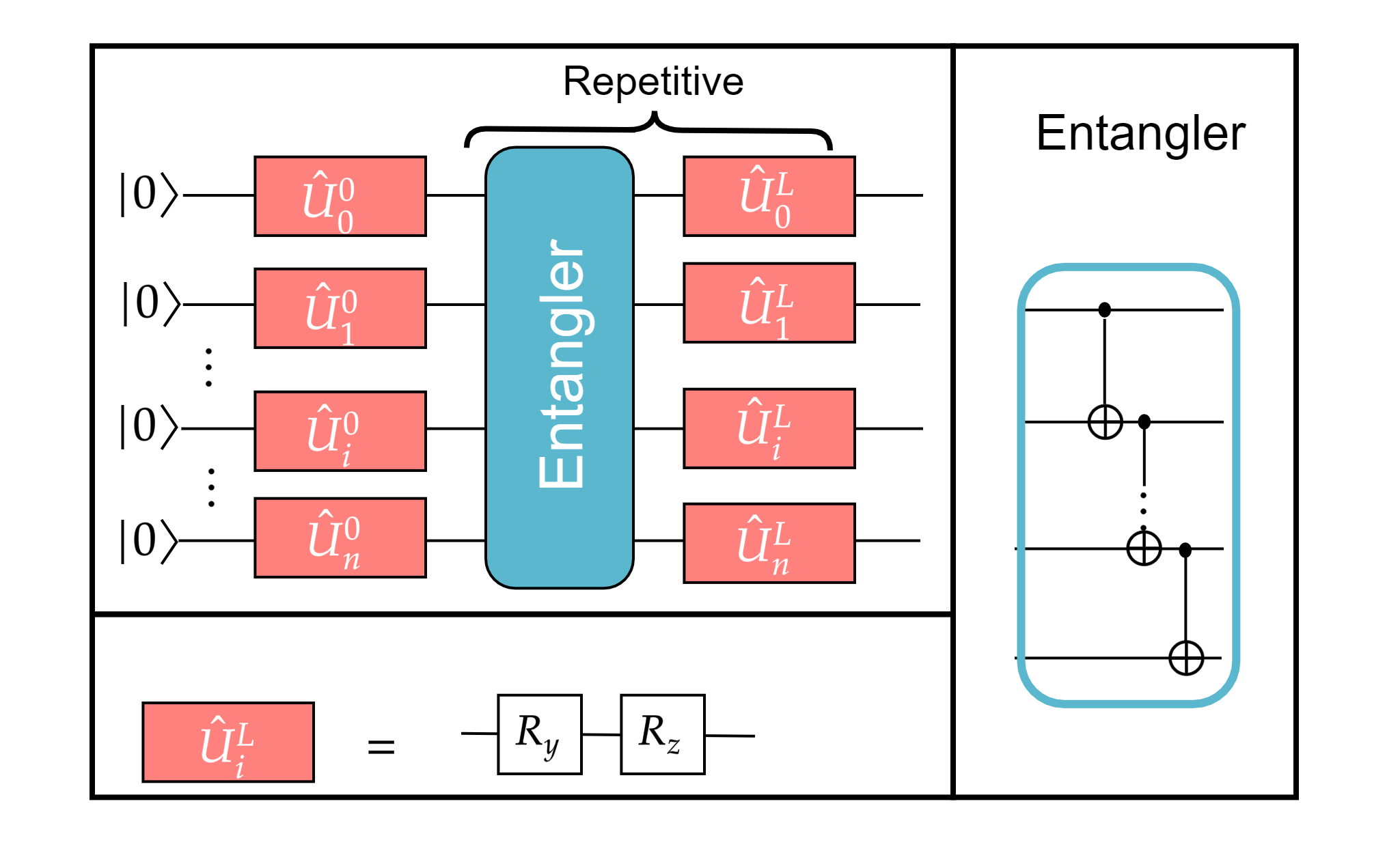}
          \caption{The circuit of Hardware Efficient ansatz. The ansatz is constructed by employing single-qubit rotations and entanglers. $\hat{U}^L_i$ is the single-qubit rotation for i-$th$ qubit in L-$th$ layer. The entangler here is chosen as a series of CNOT gates to connect the nearest-neighbor qubits and the sequence of single-qubit rotation gates are set as  $R_y, R_z$ gates.}
          \label{heapic}
      \end{figure}
      
      The structure of the HEA circuit used in this study is shown in Figure~\ref{heapic}, where every single-qubit rotation included one Ry and Rz rotations with independent parameters, respectively. 
      The entangler is chosen as a series of CNOT gates to connect the nearest-neighbor qubits
      
      To make the results of HEA more accurate, some strategies should be employed in this algorithm. For example,
      different initial values for parameters should be considered to avoid the local minimum issue\cite{Kandala2017}. 
      Following the strategy, to construct the ansatz, we started from a single layer circuit for each case of bond length. If this circuit couldn't reach chemical accuracy, we would iteratively increase the number of layers in the circuit until the maximum times of energy evaluations reach N. For each iteration, we choose the best result among S times of optimization executed with different initial variational parameters. In this study, we set N = 50000, S = 10.
    
    \subsection{k Unitary Pair Coupled Cluster with Generlized Singles and Doubles Product Wave Functions (k-UpCCGSD) }
        $k$ Unitary Pair Coupled Cluster with Generalized Singles and Doubles Product Wave Functions(k-UpCCGSD) ansatz\cite{Lee2019} depends on k products of paired coupled-cluster double excitation operators (pCCD)\cite{Stein2014} and generalized single excitation operators. "Generalized" means that spin orbitals are not distinguished between occupied and virtual ones. With very sparse double excitation operators, k-UpCCGSD gives a lower cost of quantum operations than 
        either UCCSD or UCCGSD, and can be written as follow:
        \begin{equation}
            |\psi\rangle = \prod^{k}_{\gamma=1} \left(e^{\hat{T}^{(\gamma)}-\hat{T}^{\dagger(\gamma)}}\right) \left|\Psi_{0}\right\rangle,
        \end{equation}
        and the excitation operator $\hat{T}^{(\gamma)}$ for each integer $\gamma$ includes generalized single excitation operators and pCCD excitation operators: 
        \begin{equation}
            \begin{aligned}
            &\hat{T}_1=\sum_{p q} t_{p}^{q} \hat{a}_{q}^{\dagger} \hat{a}_{p},\\
            &\hat{T}_2=\sum_{P Q} t_{P_{\alpha} P_{\beta}}^{Q_{a} Q_{\beta}} a_{Q_{\alpha}}^{\dagger} \hat{a}_{Q_{\beta}}^{\dagger} \hat{a}_{P_{\beta}} \hat{a}_{P_{\alpha}},
            \end{aligned}
        \end{equation}
        where $p$ and $q$ represent arbitrary spin orbitals, $P$ and $Q$ stand for arbitrary spatial orbitals, and $\alpha$($\beta$) is for spin-up electrons (spin-down electrons). $\hat{T}_2$ term is very sparse because it only excites electron pairs from one spatial orbital to another. 
    
    \subsection{Qubit Unitary Coupled-Cluster (QUCC) Ansatz }
        Qubit Unitary Coupled-Cluster singles and doubles ansatz(QUCCSD) \cite{Yordanov2020, Yordanov2021} is a variant of UCCSD ansatz which uses qubit excitation operators instead of Fermion excitation operators. The number of CNOT gates 
        in QUCCSD is much smaller than that in the original version of UCCSD. Single and double qubit excitation operators in Equation~\ref{qubit_ex} are generated by the qubit annihilation operators and creation operators, denoted as $Q$ and $Q^{\dagger}$, respectively:
        \begin{equation}
            \begin{aligned}
            & \hat{\tau}_{i}^{k}(\theta) = \theta\left(\hat{Q}_{k}^{\dagger} \hat{Q}_{i}-\hat{Q}_{i}^{\dagger} \hat{Q}_{k}\right),\\
            & \hat{\tau}_{i j}^{k l}(\theta) = \theta\left(\hat{Q}_{k}^{\dagger} \hat{Q}_{l}^{\dagger} \hat{Q}_{i} \hat{Q}_{j}-\hat{Q}_{i}^{\dagger} \hat{Q}_{j}^{\dagger} \hat{Q}_{k} \hat{Q}_{l}\right),
            \label{qubit_ex}
            \end{aligned}
        \end{equation}
        where
        $$
        \begin{array}{l}
        \hat{Q}^{\dagger}=|1\rangle\langle 0|=1/2(\sigma^x-i \sigma^y), \\
        \hat{Q} =|0\rangle\langle 1|=1/2(\sigma^x+i \sigma^y).
        \end{array}
        $$
        $Q^\dagger$ and $Q$ are similar to fermion operators transformed by Jordan-Wigner mapping, while without the Pauli Z chain operators (see Equation~\ref{jw} in subsection~\ref{ssec:mapping}).
     
    \subsection{Adaptive Derivative-Assembled Pseudo-Trotter (ADAPT)} \label{ADAPT VQE}
     Adaptive Derivative-Assembled Pseudo-Trotter (ADAPT) VQE \cite{Grimsley} is constructed by screening operators from a predefined pool for circuit depth reduction and accuracy improving. It grows the ansatz by introducing one operator with
     maximum gradient in the pool. 
     The circuit construction terminates when the norm of the gradient vector is lower than $\varepsilon$.
     
     \begin{figure}[htbp]
        \centering
        \includegraphics[width=0.9\linewidth]{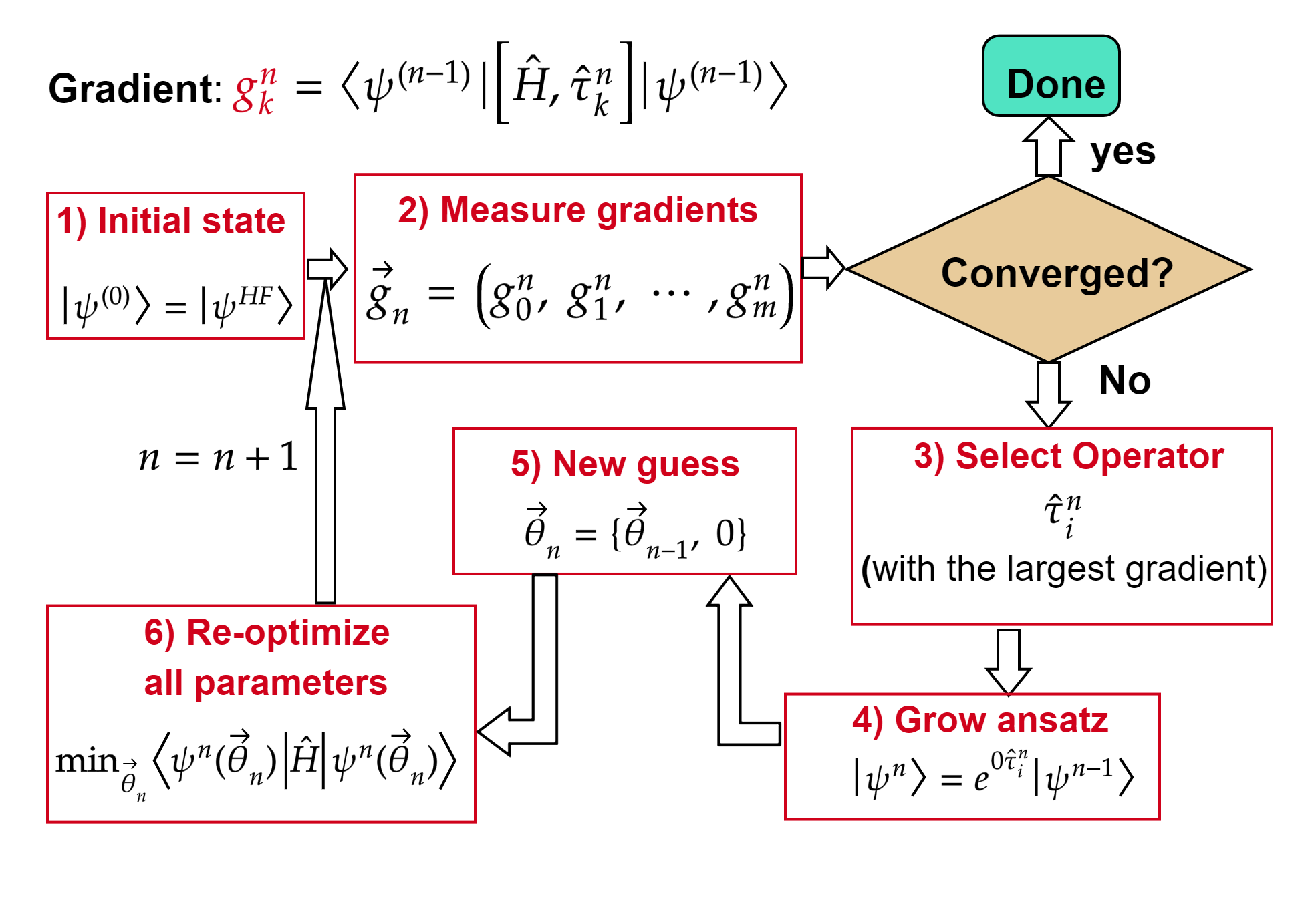}
        \caption{The flow chart of ADAPT-VQE. The initial state $\ket{\psi^{(0)}}$ is the Hartree-Fock state. Here we assume that there are m operators in the pool and $\hat{\tau}_k^n$ represents the $k$-th pool operator at $n$-th iteration. Each iteration the operator with the largest gradient is added into the ADAPT ansatz.Then all parameters are re-optimized to find a better result. The algorithm terminates when the norm of the gradient vector $\vec{g}_n$ is lower than a given threshold.}
        \label{ADAPT-VQE Flow}
    \end{figure}    
     
    The operator pool $P$ is composed of a set of spin-adapted single and double fermionic operators $\hat{\tau}$. For the $n$-th step, the gradient for $\hat{\tau}_i$ can be written as
    \begin{equation}
        g_i^n = \bra{\psi_{n-1}}[\hat{H}, \hat{\tau}_i^n]\ket{\psi_{n-1}},
        \label{gradient}
    \end{equation}
    and the corresponding gradient vector is defined as
    \begin{equation}
        \vec{g}_n = (g_0^n, \cdots, g_i^n, \cdots).
        \label{gradient-vector}
    \end{equation}
    
    The circuit construction terminates when $||\vec{g}_n||<\varepsilon$. 
    Steps of the ansatz construction are illustrated in Figure~\ref{ADAPT-VQE Flow}. 
    
    We calculated the gradient of each operator in step 2 by finite difference, $\delta$ is set as $10^{-5}$ by default, and $\varepsilon$ is $10^{-2}$. Operators in the pool are approximated with a single trotter step, and the corresponding quantum circuits are obtained after Jordan-Wigner mapping.
    
    \subsection{Qubit Coupled Cluster (QCC) Ansatz} \label{QCC}
    Qubit coupled cluster wavefunction \cite{Ryabinkin2018}  has the following form:
    \begin{equation}
    \ket{\Psi(\mathbf{\tau}, \mathbf{\Omega})} = \hat{U}(\mathbf{\tau})\ket{\mathbf{\Omega}},
    \end{equation}
    where $\ket{\mathbf{\Omega}}$ represents the mean-field part while $\hat{U}(\mathbf{\tau})$ introduces correlation parts and $\tau$ is a set of variational parameters.
    The mean-field wavefunction is a product of single-qubit coherent states.
    \begin{equation}
        \ket{\mathbf{\Omega}} = \prod_i\ket{\Omega_i}
    \end{equation}
    \begin{equation}
        \ket{\Omega_i}=\cos{(\frac{\theta_i}{2})}\ket{0}+e^{i\phi_i}\sin{(\frac{\theta_i}{2})}\ket{1}
        \label{Bloch state}
    \end{equation}
    where Equation~\ref{Bloch state} is a "Bloch state" for qubit $i$. The correlation is introduced by entanglers $e^{i\tau_i\hat{p}_i}$, and 
    \begin{equation}
    \hat{U}(\mathbf{\tau}) = \prod_k e^{i\tau_k\hat{p}_k}.
    \end{equation}
    where $\hat{p}_k$ are the Pauli strings. 
    
    QCC wavefunction requires $2N_q+N_{ent}$ independent parameters, where $2N_q$ variational parameters belong to the mean-field part and $N_{ent}$ is the number of entanglers. In the project, we benchmarked an iterative modification of the QCC scheme \cite{Ryabinkin2020}. In this scheme, each entangler is ranked based on the value of $\Delta E[\hat{p}_k]$ as
    \begin{equation}
        \Delta E[\hat{p}_k] = \min_{\tau}E[\tau, \hat{p}_k] - E[0, \hat{p}_k],
    \end{equation}
    where $E[0, \hat{p}_k]=E_{QMF}$ is the qubit mean-field energy. All entanglers are ranked by their contribution to energy reduction compared to $E_{QMF}$. After that, fix the entangler with the best performance as a layer of the ansatz. The procedure will be iteratively done for more layers until an energy threshold $\epsilon = 0.0016$ Hartree (chemical accuracy) is reached.
    
    \subsection{qubit-ADAPT} \label{qubit-ADAPT VQE}
    Different from ADAPT-VQE, qubit-ADAPT VQE \cite{Tang2019} constructs ansatz with Pauli strings 
    rather than fermionic operators. 
    Such method reduces the circuit depths while increasing the construction iterations.
    
    To generate the "qubit pool" in the qubit-ADAPT VQE ansatz, fermionic operators are transformed by Jordan-Wigner mapping  and divided into a pile of individual Pauli strings, $\hat{\tau}=\otimes\prod_i \hat{p}_i$, where $\hat{p}_i \in \{\hat{\sigma}^I, \hat{\sigma}^x, \hat{\sigma}^y, \hat{\sigma}^z\}$. These Pauli strings contain an odd number of Y's. Numerical simulation shows that the Pauli Z chains can be eliminated to reduce circuit depth further. Gradients of those operators are given by
    \begin{equation}
        \frac{\partial \langle E \rangle}{\partial \theta} = i\langle\psi|[\hat{\tau}, \hat{H}]|\psi \rangle.
    \end{equation}
    Steps of the algorithm can also refer to Figure~\ref{ADAPT-VQE Flow}.
    
    The parameter shift rule\cite{Mitarai2018} is used to calculate gradients of operators like $e^{i\theta\hat{\tau}}$, where $\hat{\tau}$ represents a single Pauli string. The "qubit pool" is obtained in the ADAPT-VQE with Jordan-Wigner mapping and Pauli Z operators eliminated.

    \subsection{Low-depth Circuit Ansatz (LDCA)} \label{LDCA VQE}
    Motivated by the Bogoliubov coupled cluster theory, \cite{Dallaire_Demers_2019} presented a new type of low-depth circuit ansatz(LDCA). Using the theory of matchgates, it showed how pure fermionic Gaussian states can be exactly prepared on a quantum computer using a linear-depth circuit.
    
\begin{figure*}
        \centering
        \includegraphics[width=\textwidth]{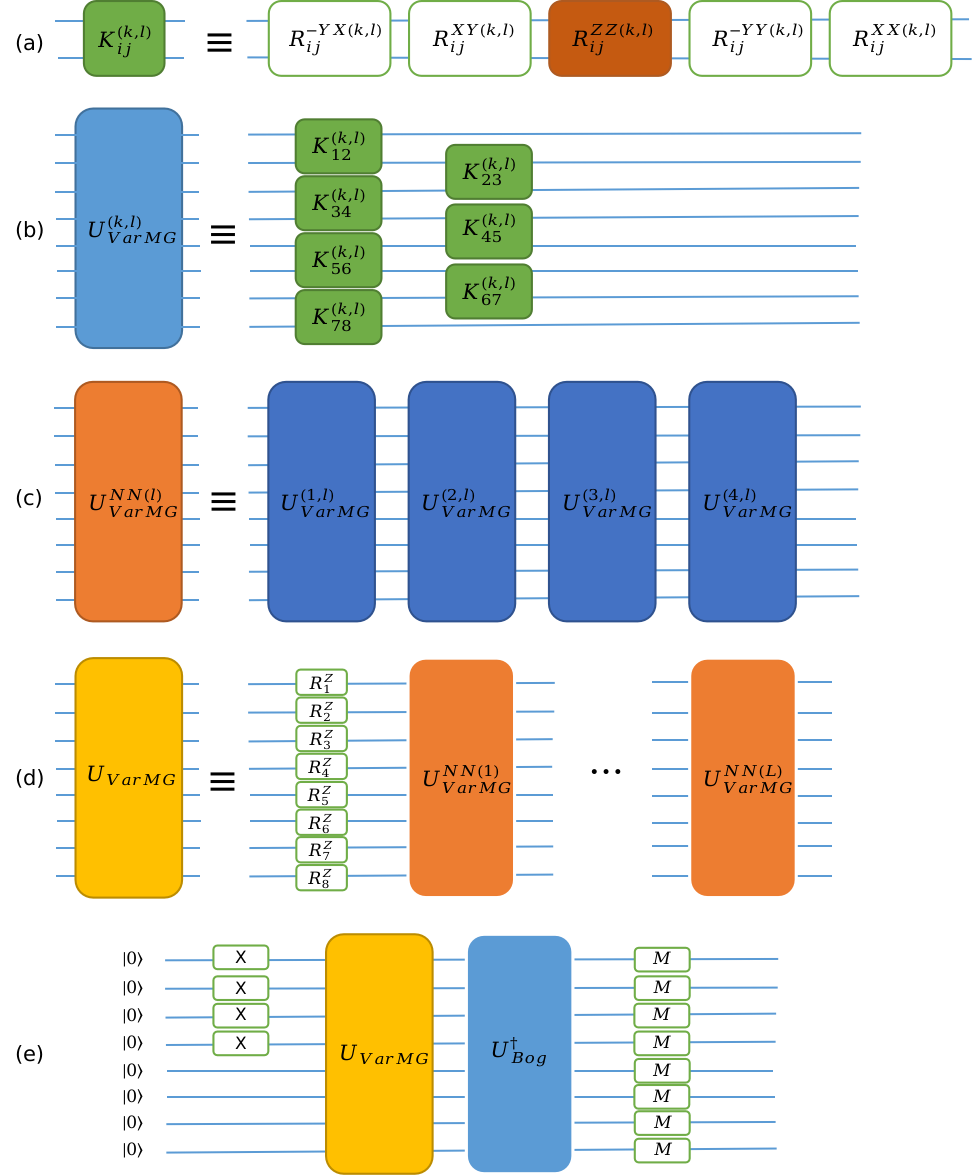}
        \caption{Gate decomposition of the L-cycle LDCA on a linear chain of $M=8$ qubits.
        (a) The matchgate $K_{ij}^{(k, l)}$ is a 2-local operation between qubits $i$ and $j$ for layer $k$ cycle $l$. (b) The $U_{VarMG}^{(k, l)}$ for layer $k$ cycle $l$ is built by operating $K_{ij}^{(k, l)}$ first on the even pairs of qubits and then on the odd pairs. (c) The circuit of cycle l $U_{VarMG}^{NN(l)}$ is composed by $\lceil \frac{M}{2} \rceil$ layers. (d) The L-cycle construction of $U_{VarMG}$ with one round of variational phase rotations. (e) The full LDCA protocol with initial state in Hartree-Fock and the transformation to the original fermionic basis using $U_{Bog}^\dagger$~\cite{Dallaire_Demers_2019}.}
        \label{LDCA}
    \end{figure*}

    In LDCA ansatz we use a series of 2-qubit rotations to build matchgates. Each of those rotations has a parameter to be optimized. The decomposition of Bogoliubov transformation in the LDCA can be found in \cite{Dallaire_Demers_2019}. The full LDCA protocol is shown in Figure~\ref{LDCA}.
    
    In this work, we use $L=2$ cycles of $U_{VarMG}$ as the original article shows, and considering the device limitation. The initial state is Hartree-Fock state and initial parameters are generated by random sampling in $[0, 2\pi]$. We sampled 20 times and chose the best parameters which could achieve the lowest energy.

    \subsection{Basis Rotation Circuit (BRC) Ansatz} \label{BRC VQE}
    Basis rotation circuit(BRC) ansatz realizes the Givens rotation approach to noninteracting fermion evolution \cite{AIQuantum2020}. The ansatz can be formulated as 
    \begin{equation}
        |\psi_\kappa \rangle = {U}_\kappa|\eta \rangle
    \end{equation}
    where $|\eta \rangle=a_{\eta}^{\dagger}...a_1^{\dagger}|0 \rangle$ is a computational basis state in the core orbital basis. ${U}_\kappa$ is a unitary given by
    \begin{equation}
        {U}_\kappa = {\rm exp}(\sum_{p,q=1}^N \kappa_{pq} a_p^{\dagger} a_q)
    \end{equation}
    where $\kappa$ is an N×N anti-Hermitian matrix and $\kappa_{pq}$ is the $p$,$q$ element of $\kappa$. By taking $a_p^{\dagger}$ and $a_q$ to be fermionic creation and annihilation operators for the core orbital $\phi_{p}(r)$, we can implement variational relaxation of the $\kappa$ parameters to minimize the energy of $|\psi_\kappa \rangle$.
    
\begin{figure*}
        \centering
        \subfigure[]{\includegraphics[width=0.8\textwidth]{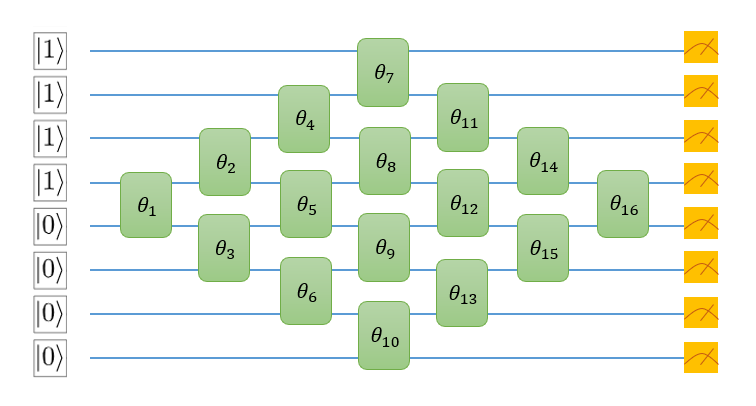}}
        \subfigure[]{\includegraphics[width=0.9\textwidth]{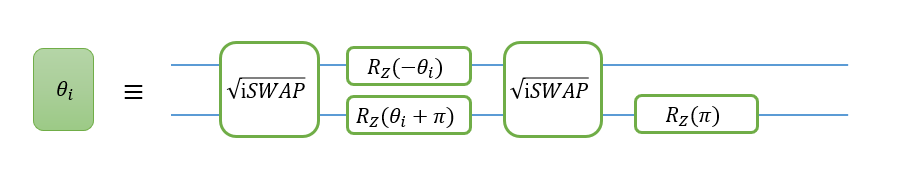}}
         \caption{$\textbf{Basis rotation circuit and compilation.}$ The figure depicts the basis rotation ansatz for a linear chain of four hydrogen atoms. Each white box with a rotation angle $\theta$ represents a Givens rotation gate which can be compiled using two $\sqrt{iSWAP}$ gates and three single-qubit Z rotation gates.}
         \label{BRC}
        \end{figure*}

    The initial state of the ansatz was obtained by filling the $\eta$ lowest energy orbitals (HF state), where $\eta$ is the number of electrons. Utilizing the optimal compilation of \cite{PhysRevLett.120.110501}, ${U}_\kappa$ can be compiled exactly (without Trotterization) into a linearly connected architecture as shown in Figure~\ref{BRC}. The BRC ansatz contains $\eta$(N-$\eta$) Givens rotations. Each Givens rotation contains one variational parameter. In this work, the initial parameters were set to the parameters obtained by randomly sampling from a uniform distribution $[-\pi, \pi)$. For each optimization, we sampled 20 times and chose the parameters that gives the lowest energy solution. 
    
\section{Benchmarking Toolkit} \label{sec:tools}
  In this section, we will introduce the benchmarking toolkit used in the study, including the data recording and comparison codes. Researchers can use these open source codes to conveniently add more benchmarking results.
  
  \subsection{Data Recording}
    In the MindQuantum package, all the original benchmarking data are recorded in the \blue{'data'} directory as \blue{.json} files. For researchers who want to benchmark more results, they may record the data in the files as the format of the existed results. For example, if the researchers want to add PES results of some molecule calculated by other ansatzes, they may find the corresponding file to record energy of the molecule, and add an item in the \blue{'energies'} dictionary. The item should be a list looks like
    \makebox[\textwidth][c] {\red{'UCCSD': [ ]}}, 
    and the energies of the molecule at different bond lengths that calculated by this ansatz should be added in the list.
    
    To make it convenient to record the data in the corresponding \blue{.json} files. One can use the functions in \blue{data.py}. The file including functions \red{initdata(), savedata() and rounddata()}, which can be used to generate the \blue{.json} file of the molecule and manage the data in it.
    
  \subsection{Data Comparison}
    Using the data in \blue{.json} files, one can compare the data and generate the pictures as the section~\ref{sec:results} shows by \blue{drawpic.py} file. The basic comparison of energy error, runtime until convergence, and number of parameters between different ansatzes can be made by functions such as \red{draw\_errors(), draw\_runtimes(), draw\_paras()}.
    
    To compare results calculated by different ansatzes, in \blue{drawpic.py}, one should firstly specify the marker and color of the ansatzes at the beginning of the file. Then, the results can be compared by calling the functions in the file. For example, to generate the figures that make comparison between different ansatzes for $\mol{H_4}$ molecule, one can use the sample code below:
    
    {\fontfamily{Courier New Bold}      \footnotesize           
    \begin{lstlisting}
    molecule = "H4"
    figsize=(10, 7.5)
    ansatzes = ["CCSD", "HEA", "ADAPT", "qubit-ADAPT", "QCC"]
    draw_errors(molecule=molecule, figsize=figsize, ansatzes=ansatzes)
    draw_runtimes(molecule=molecule, figsize=figsize, ansatzes=ansatzes)
    draw_paras(molecule=molecule, figsize=figsize, ansatzes=ansatzes)
    \end{lstlisting}}

\section{Benchmarking Results}\label{sec:results}
  Using the benchmarking toolkit, we benchmarked the ansatzes introduced in Section~\ref{sec:ansatz} on six representative molecules, including 8 qubits $\mol{H_4}$, 12 qubits $\mol{LiH}$, 14 qubits $\mol{BeH_2}$, 14 qubits $\mol{H_2O}$, 18 qubits $\mol{CH_4}$, and 20 qubits $\mol{N_2}$. The Molecular Hamiltonians were calculated under the STO-3G basis and transformed by Jordan-Wigner mapping. Simulations were performed on a workstation with 88 cores CPU and 224 GB RAM. The number of threads was set as 4 in all cases and MindQuantum\cite{MindQuantum} was used as the quantum simulator.
  
  To make it readable, we only show partial results in this article. As Table~\ref{tab:ansatzes} in Chapter \ref{sec:ansatz} shows, we group the ansatzes into two categories to compare: fixed-circuit ansatzes and changeable-circuit ansatzes. The fixed-circuit ansatzes include UCCSD, UCCSD0, k-UpCCGSD, and QUCC ansatzes. The changeable-circuit ansatzes include layered ansatz HEA, LDCA, BRC and adaptive ansatzes such as ADAPT, QCC, and qubit-ADAPT. Due to the RAM limitation, QCC ansatz was only 
  benchmarked on $\mol{H_4}$; qubit-ADAPT ansatz was only 
   benchmarked on $\mol{H_4}$ and $\mol{LiH}$; ADAPT ansatz were 
    benchmarked on $\mol{H_4}$, $\mol{LiH}$ and $\mol{BeH_2}$. 
  
  \subsection{Comparison of Energy Error}
  
    This part shows the results of calculated energies. To make it clear, we also introduced energies calculated by the classical HF method, CCSD method, and FCI method. We set FCI energy as the reference and showed the energy error compared to it. Since the energy error of the HF method made other results indistinguishable in the figure, it was not included. 

\begin{figure*} [htbp]
		\centering
		\noindent\makebox[\textwidth][c] {
		\hspace{9mm}
		\includegraphics[width=0.32\paperwidth]{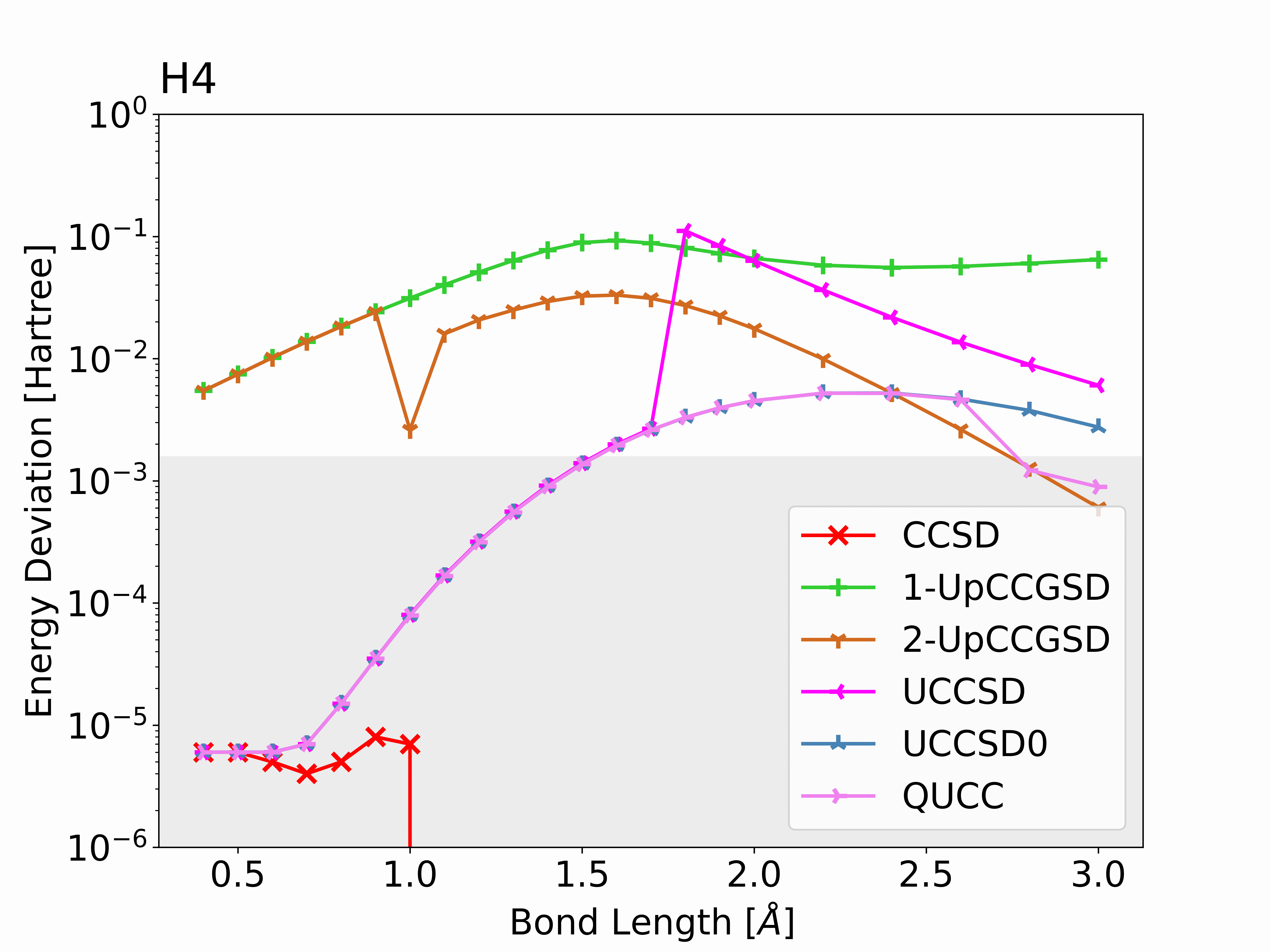}
		\hspace{-9mm}
		\includegraphics[width=0.32\paperwidth]{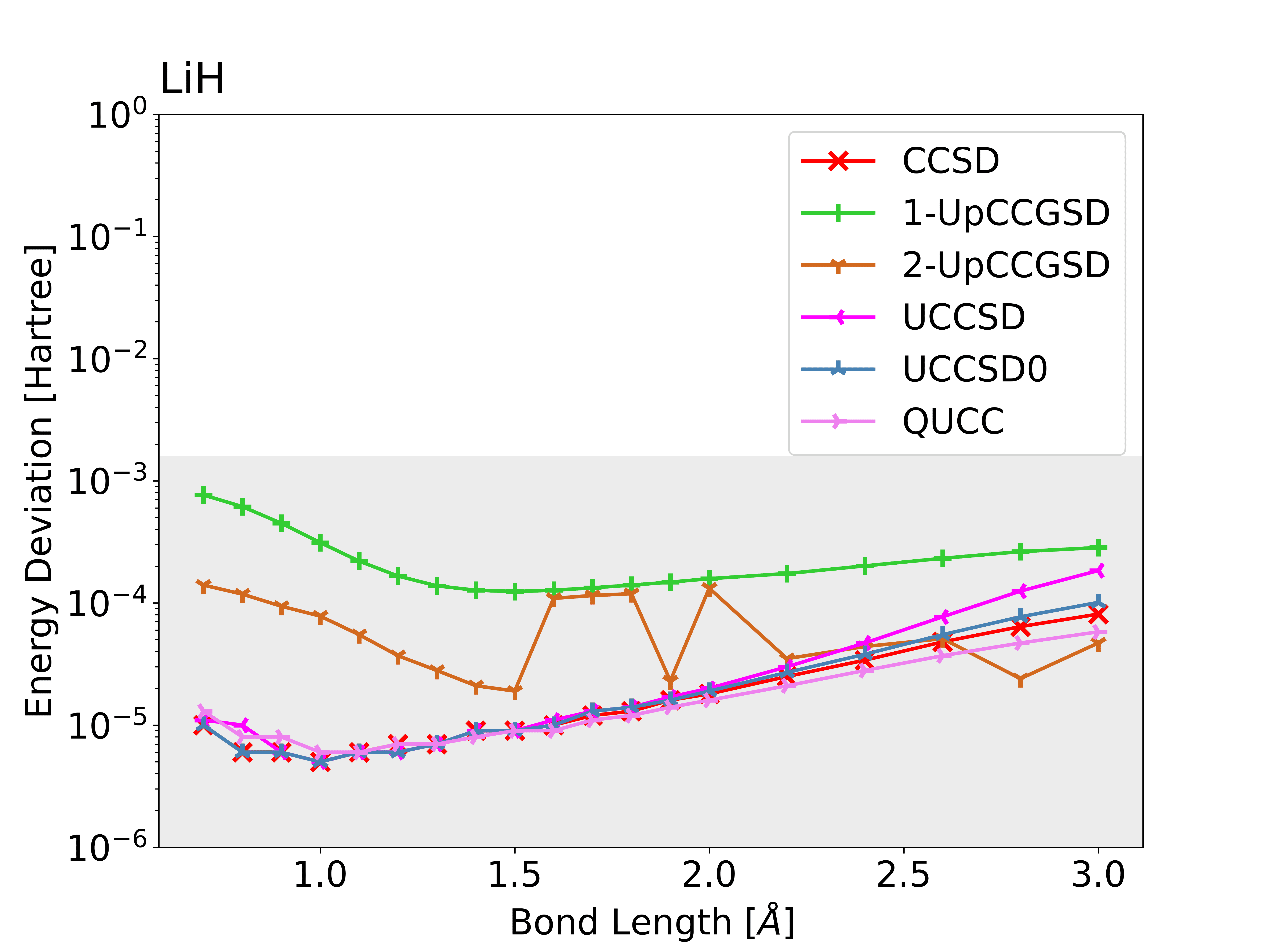}
		\hspace{-9mm}
		\includegraphics[width=0.32\paperwidth]{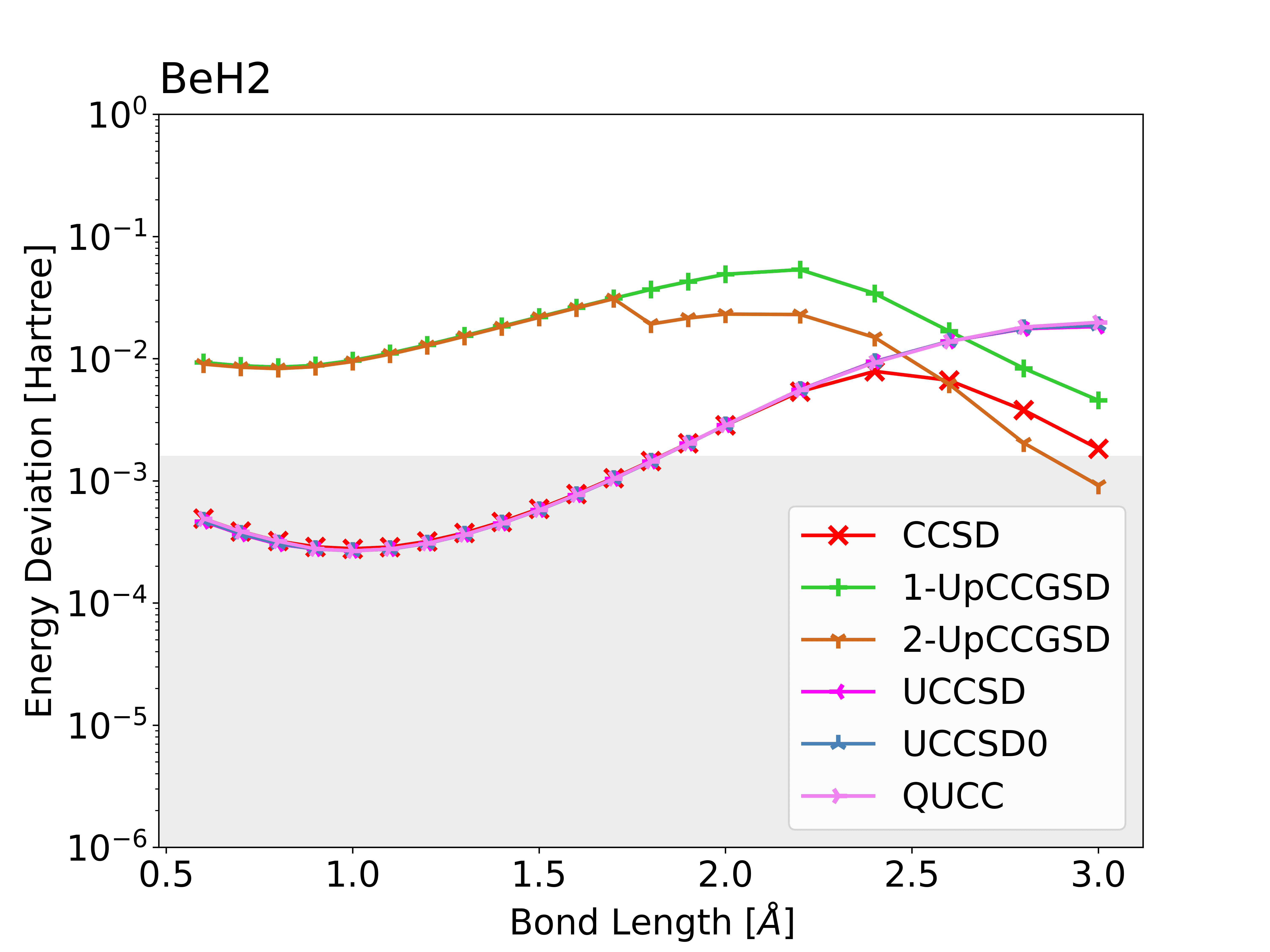}
		}
		\noindent\makebox[\textwidth][c] {
		\hspace{9mm}
		\includegraphics[width=0.32\paperwidth]{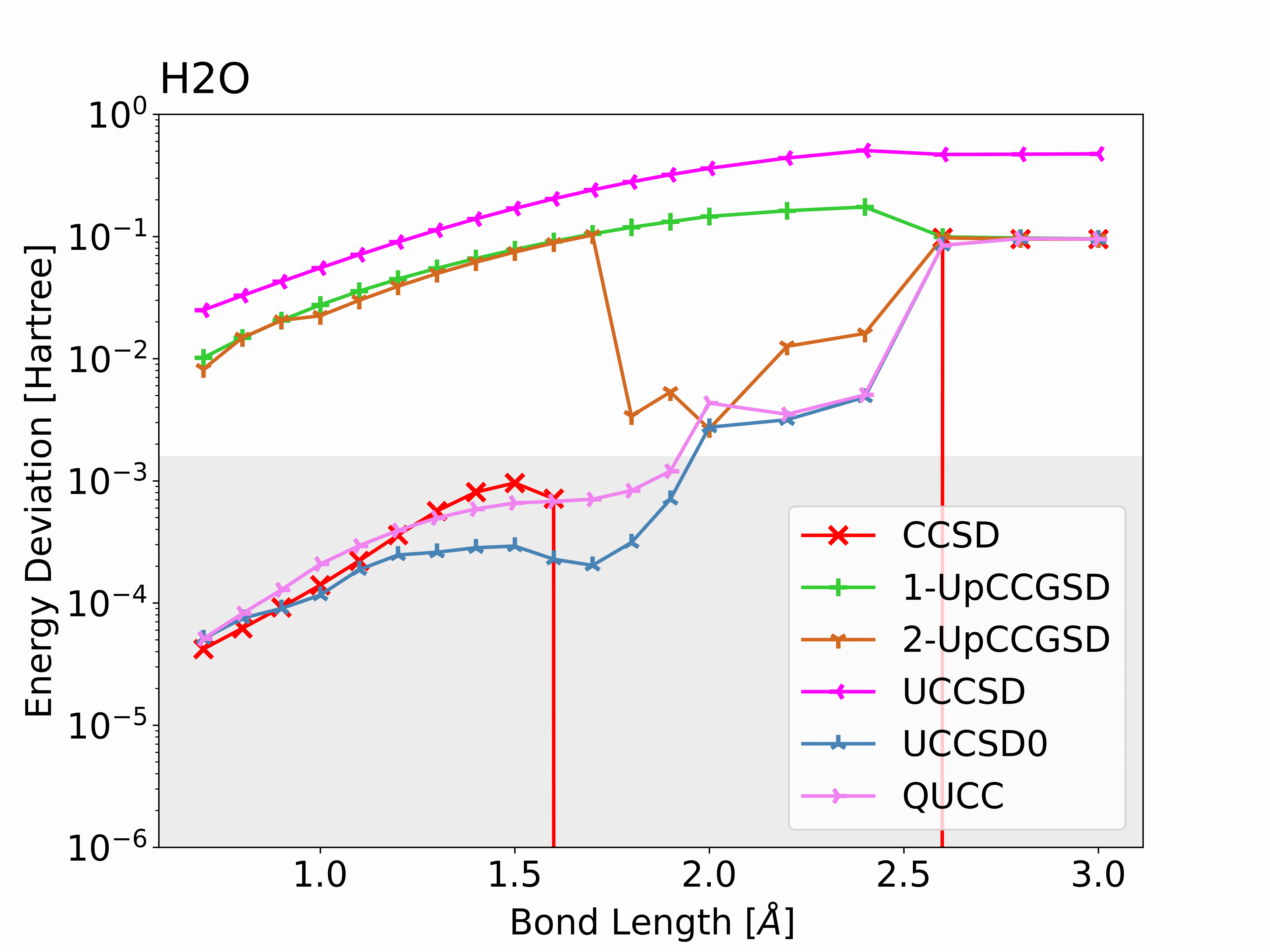}
		\hspace{-9mm}
		\includegraphics[width=0.32\paperwidth]{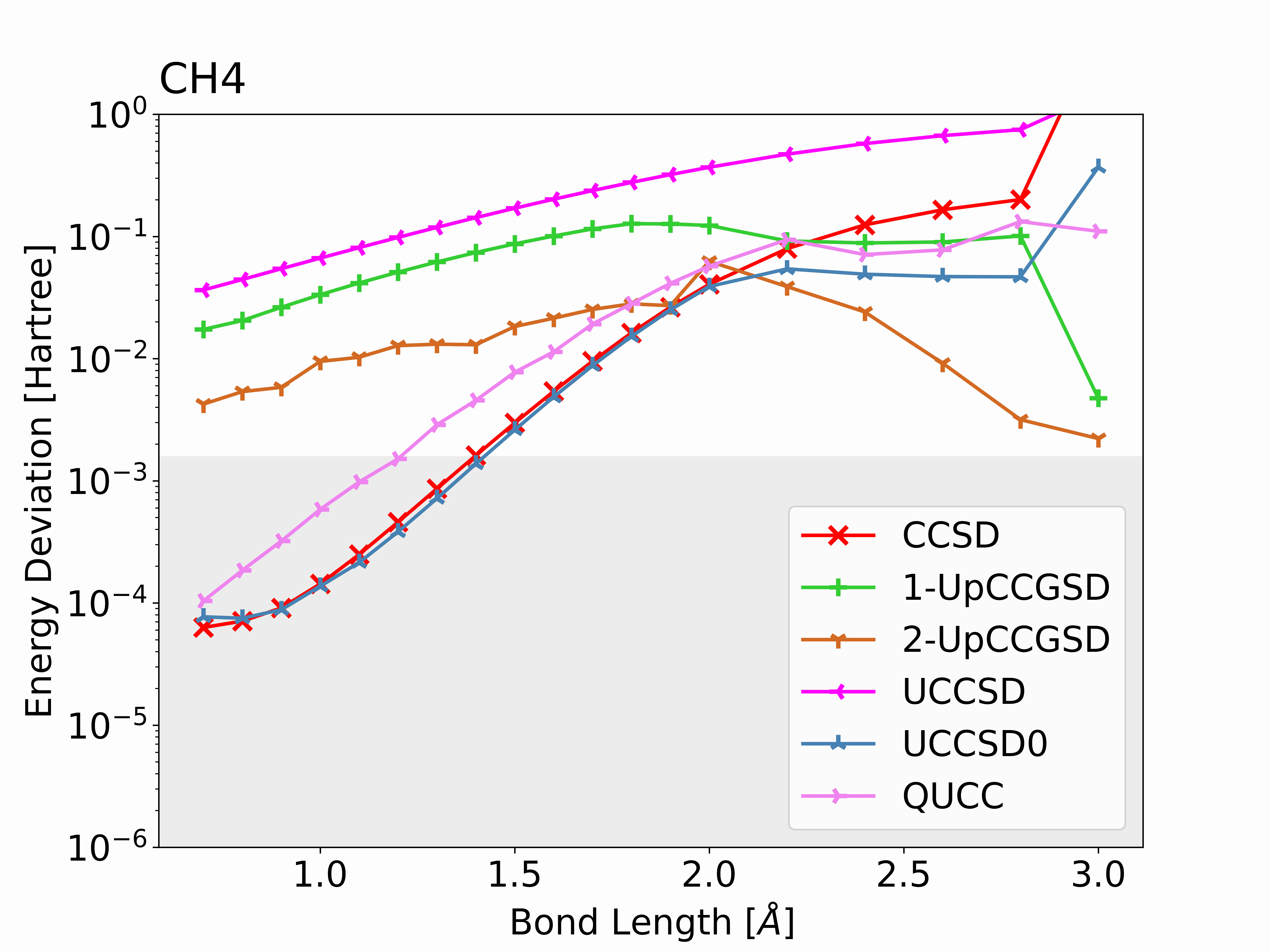}
		\hspace{-9mm}
		\includegraphics[width=0.32\paperwidth]{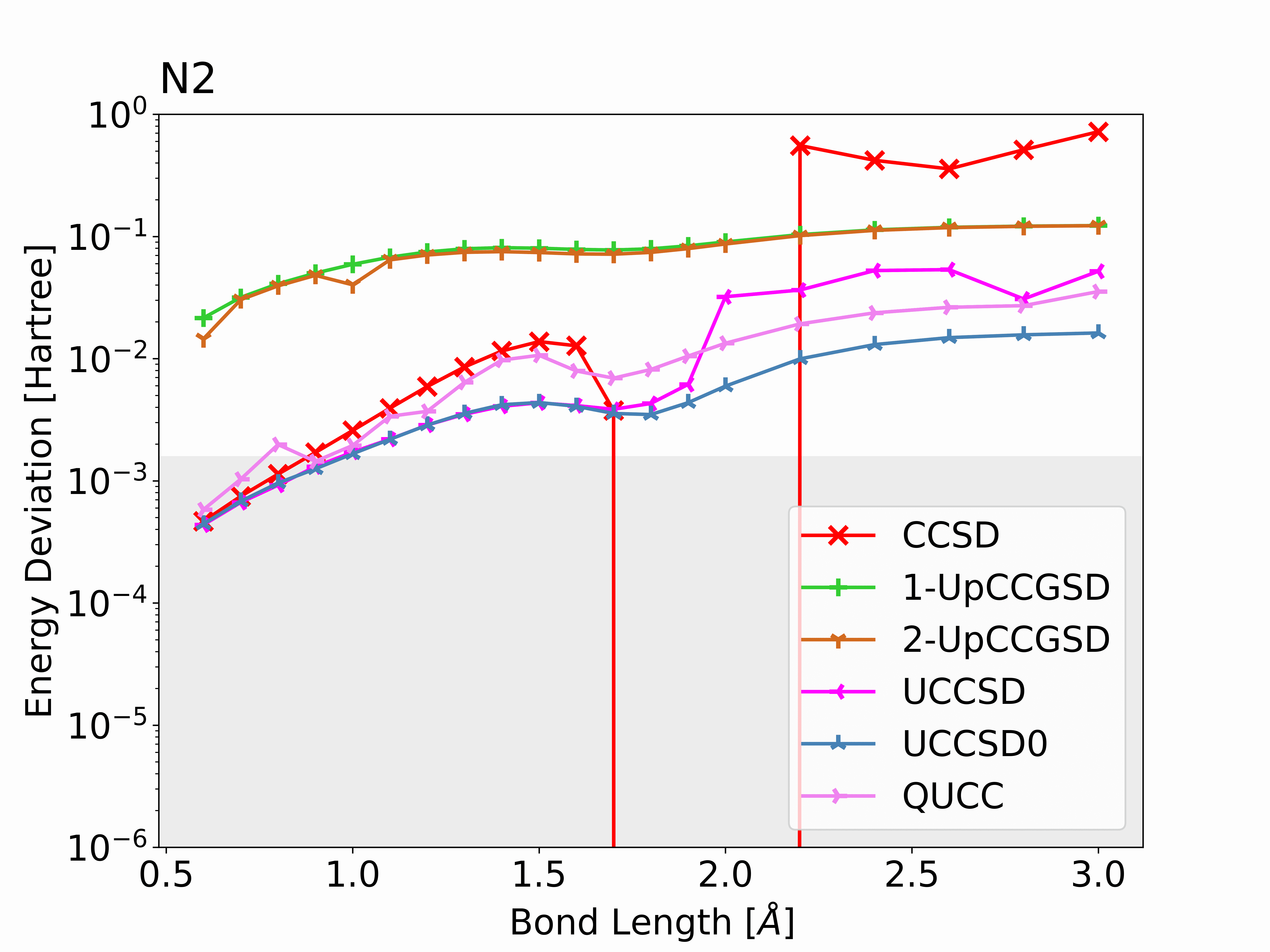}
		}
		\caption{The energy error of $\mol{H_4}, \mol{LiH}, \mol{BeH_2}, \mol{H_2O}, \mol{CH4}$,  and $\mol{N_2}$ calculated by CCSD method and fixed-circuit ansatzes compared to the full-CI method at different bond length. The results of full-CI method are settled as reference(0 energy). Some energy points calculated by CCSD are lower than full-CI energy, which can't be shown in the figure. The grey area denotes the chemical accuracy (-0.0016 $<$ energy error $<$ 0.0016).}
		\label{fig:energy1}
	\end{figure*}

    The results calculated by fixed-circuit ansatzes and the CCSD method are shown in Figure~\ref{fig:energy1}. Noticing that the CCSD energy is not bounded by variational principle and may yield lower energy than the exact ground energy in cases like stretched bond of $\mol{H_4}, \mol{H_2O}, \mol{N_2}$. From the results, UCCSD0 always behaved better than UCCSD, which means UCCSD0 owned a better parameterization way for tested molecules. The CCSD method failed to describe the ground state of molecules with a stretched bond. Compared to other ansatzes, 1-UpCCGSD and 2-UpCCGSD have relatively more errors and usually can not reach the chemical accuracy. UCCSD0 had the best performance among these ansatzes, since it's relatively accurate in most cases. QUCC ansatz also had a good performance, but for larger systems, e.g., $\mol{N_2}$, it performs worse than that of the UCCSD0 ansatz. 
    Furthermore, for the cases with bond length longer than 1.4 {\AA} in $\mol{H_4}$, all the ansatzes were challenging to approach the exact ground state energies. This phenomenon also showed in bond stretched $\mol{BeH_2}, \mol{H_2O}$, which revealed a limitation for all these ansatzes.

\begin{figure*} [htbp]
		\centering
		\noindent\makebox[\textwidth][c] {
		\hspace{9mm}
		\includegraphics[width=0.32\paperwidth]{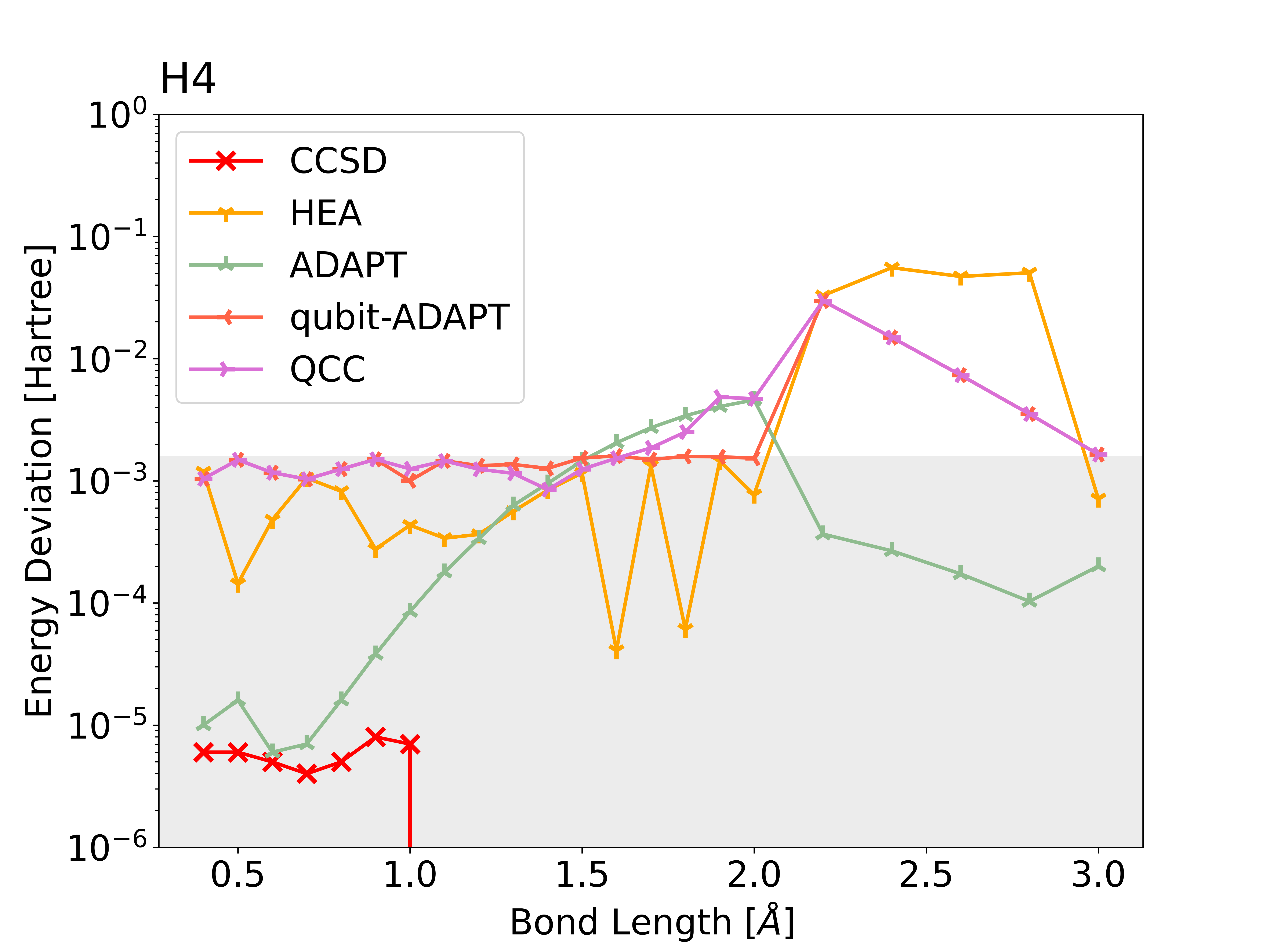}
		\hspace{-9mm}
		\includegraphics[width=0.32\paperwidth]{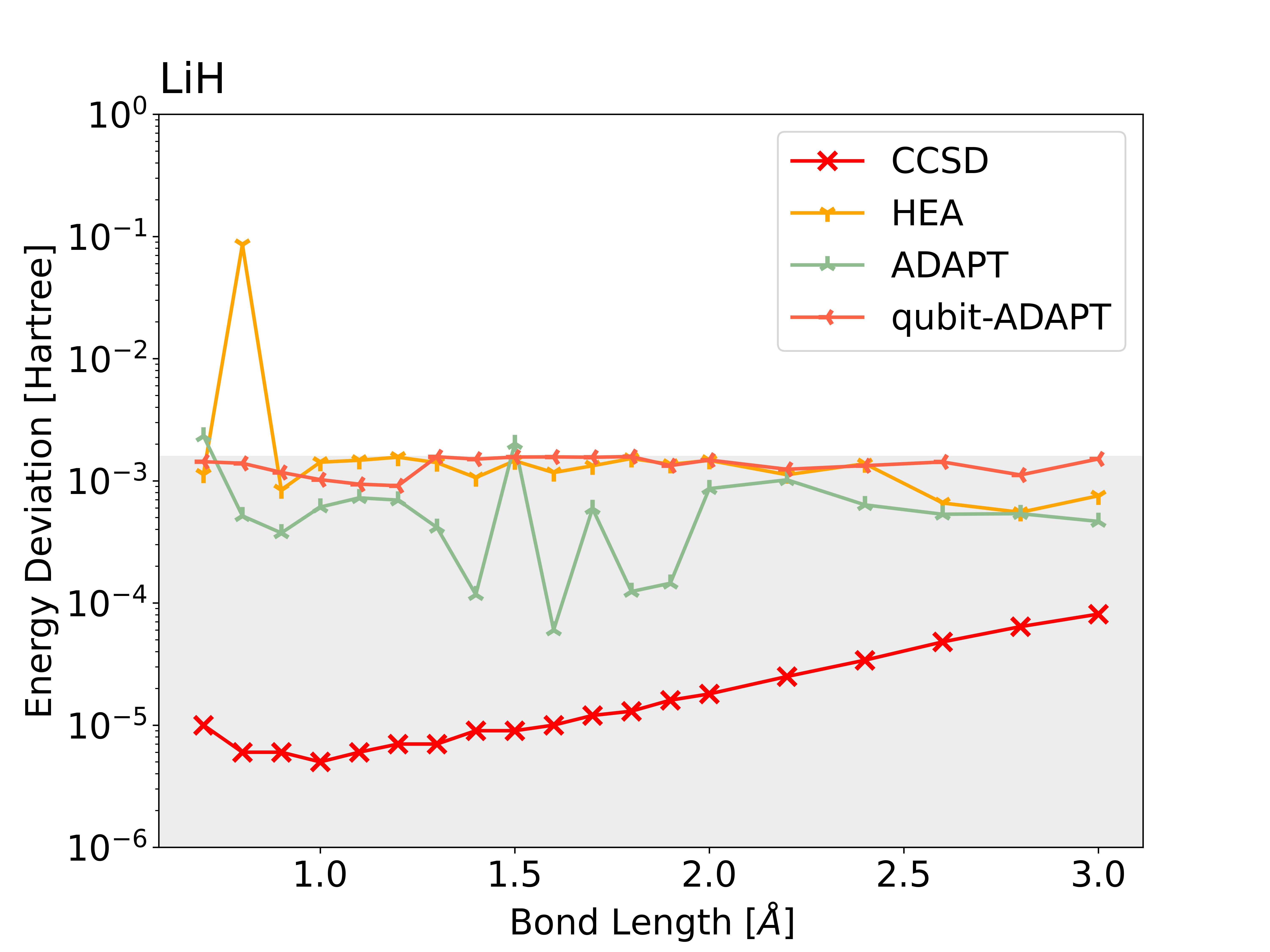}
		\hspace{-9mm}
		\includegraphics[width=0.32\paperwidth]{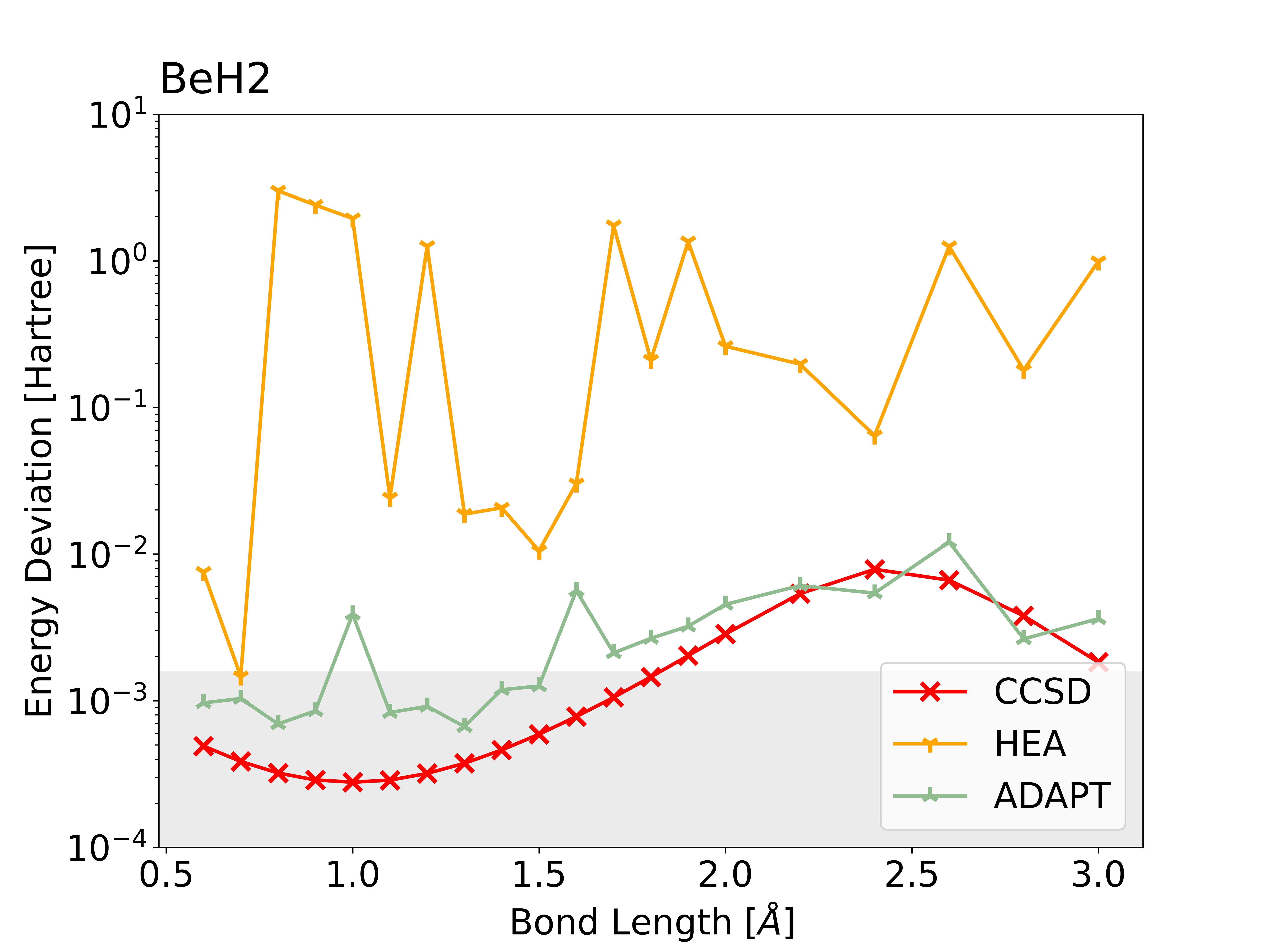}
		}
		\caption{
		The energy error of $\mol{H_4}, \mol{LiH}$ and $\mol{BeH_2}$ calculated by CCSD method and changeable-circuit ansatzes compared to the full-CI method at different bond length (Noticing that the VQE procedure will stop when maximal iteration or the chemical accuracy is reached). The results of full-CI method are settled as reference(0 energy). Some energy points calculated by CCSD are lower than full-CI energy, which can't be shown in the figure. The grey area denotes the chemical accuracy (-0.0016 $<$ energy error $<$ 0.0016). }
		\label{fig:energy2}
	\end{figure*}

	The energy results calculated by changeable-circuit ansatzes except for LDCA and BRC are shown in Figure~\ref{fig:energy2}. It can be seen that these ansatzes had big error at some specific points of bond lengths. It may because the optimizer was trapped into local minimum at some cases of bond lengths for these ansatzes, indicating that these ansatzes are not stable.

    The above results showed that UCCSD0, 2-UpCCGSD, QUCC, and ADAPT ansatzes had relatively good performance among the benchmarked ansatzes. UCCSD0 and QUCC usually obtained more accurate energy compared to the other ansatzes, but UCCSD0 was more accurate than QUCC for larger systems, such as $\mol{N_2}$ molecule. However, they still had high energy error when bond lengths were large enough. 2-UpCCGSD may have less error in some cases, especially for $\mol{CH_4}$ at stretched bond lengths. The ADAPT ansatz had a relatively good performance in smaller molecules (below 14 qubits). But it failed to run on our hardware for larger molecules. The computation resource of adaptive methods increased rapidly as the system size grew, which means the ADAPT ansatz may be computational prohibitive for larger systems. 
    
    \begin{figure*} [htbp]
		\centering
		\noindent\makebox[\textwidth][c] {
		\hspace{9mm}
		\includegraphics[width=0.32\paperwidth]{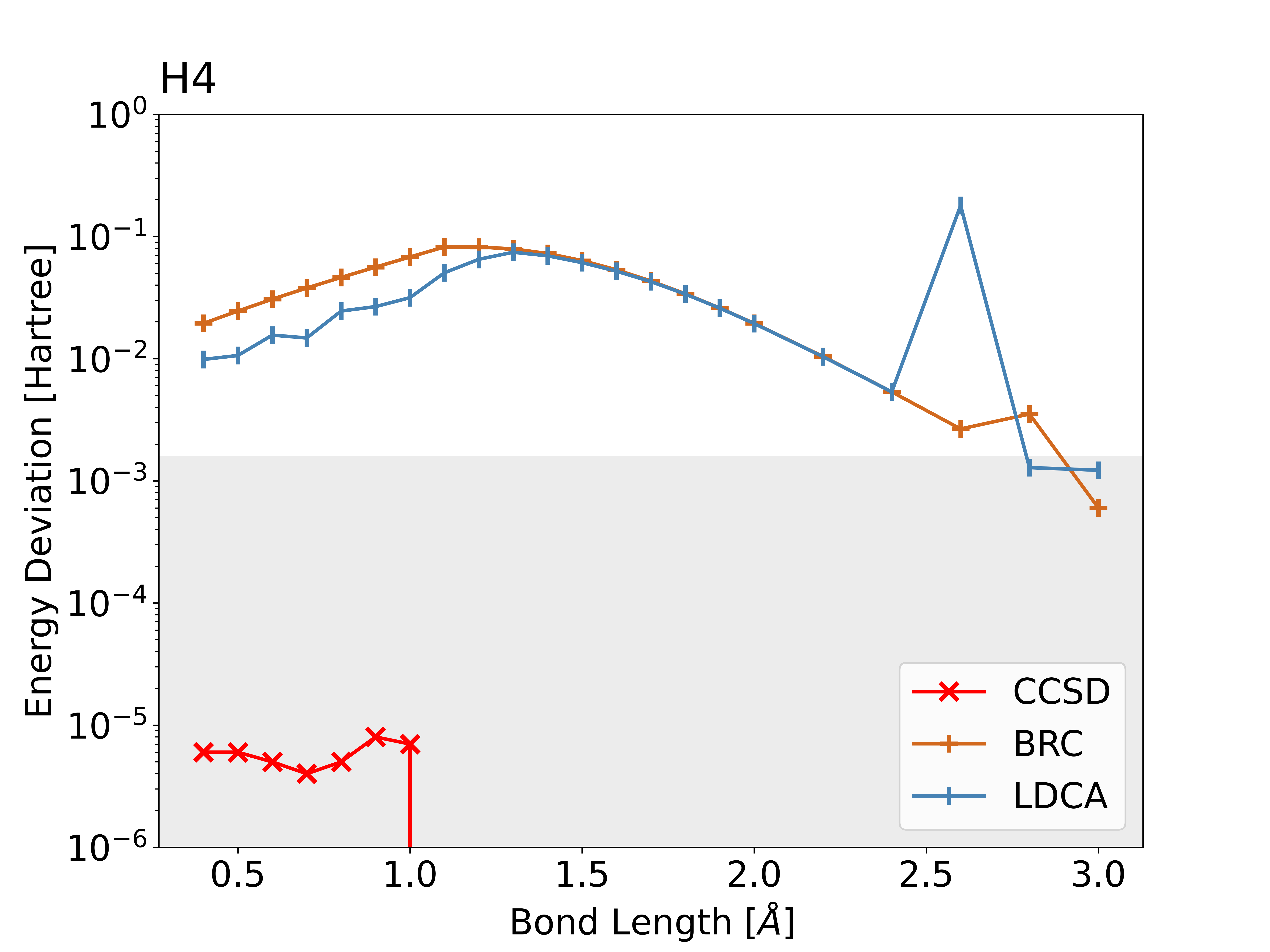}
		\hspace{-9mm}
		\includegraphics[width=0.32\paperwidth]{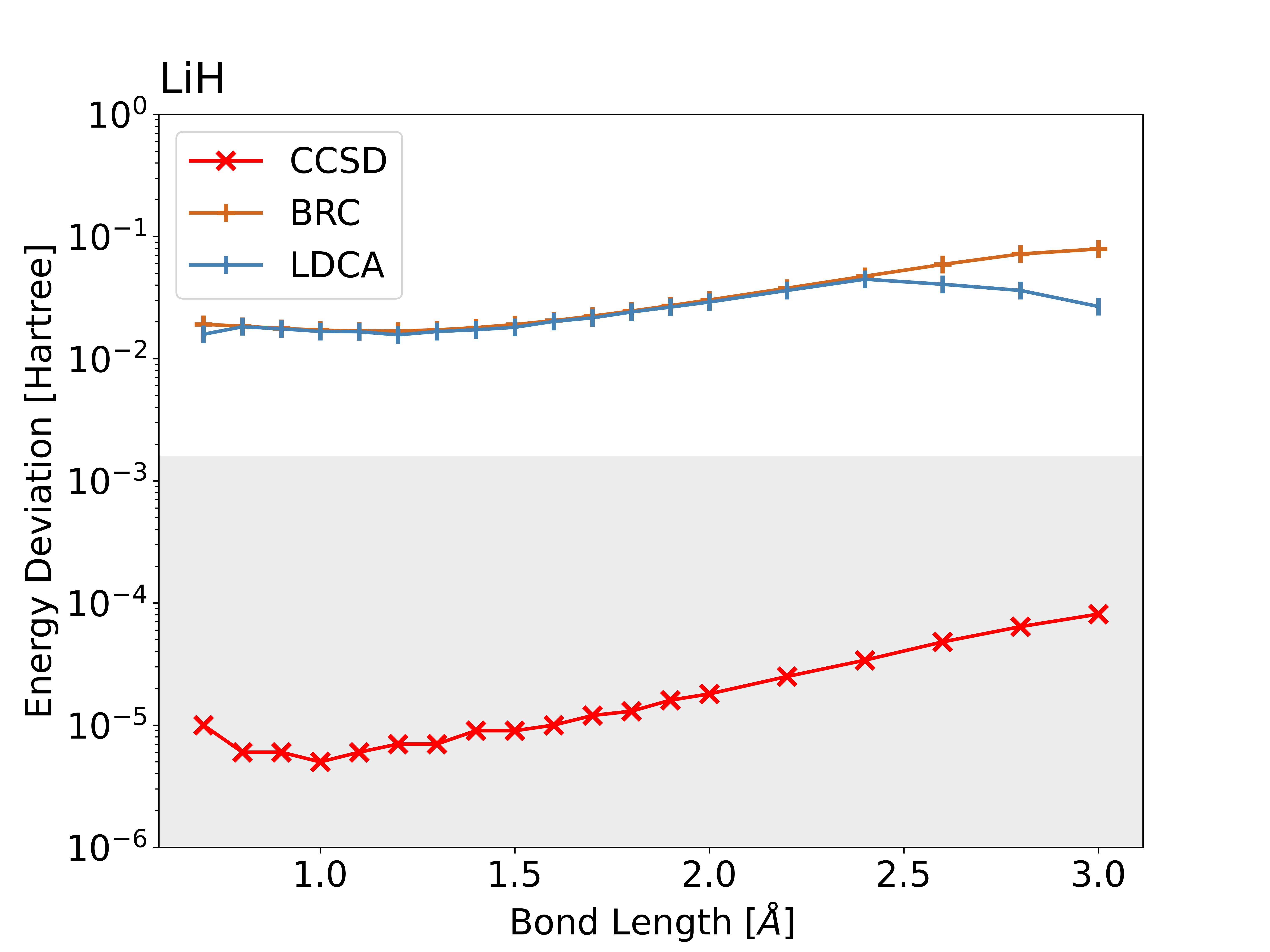}
		\hspace{-9mm}
		\includegraphics[width=0.32\paperwidth]{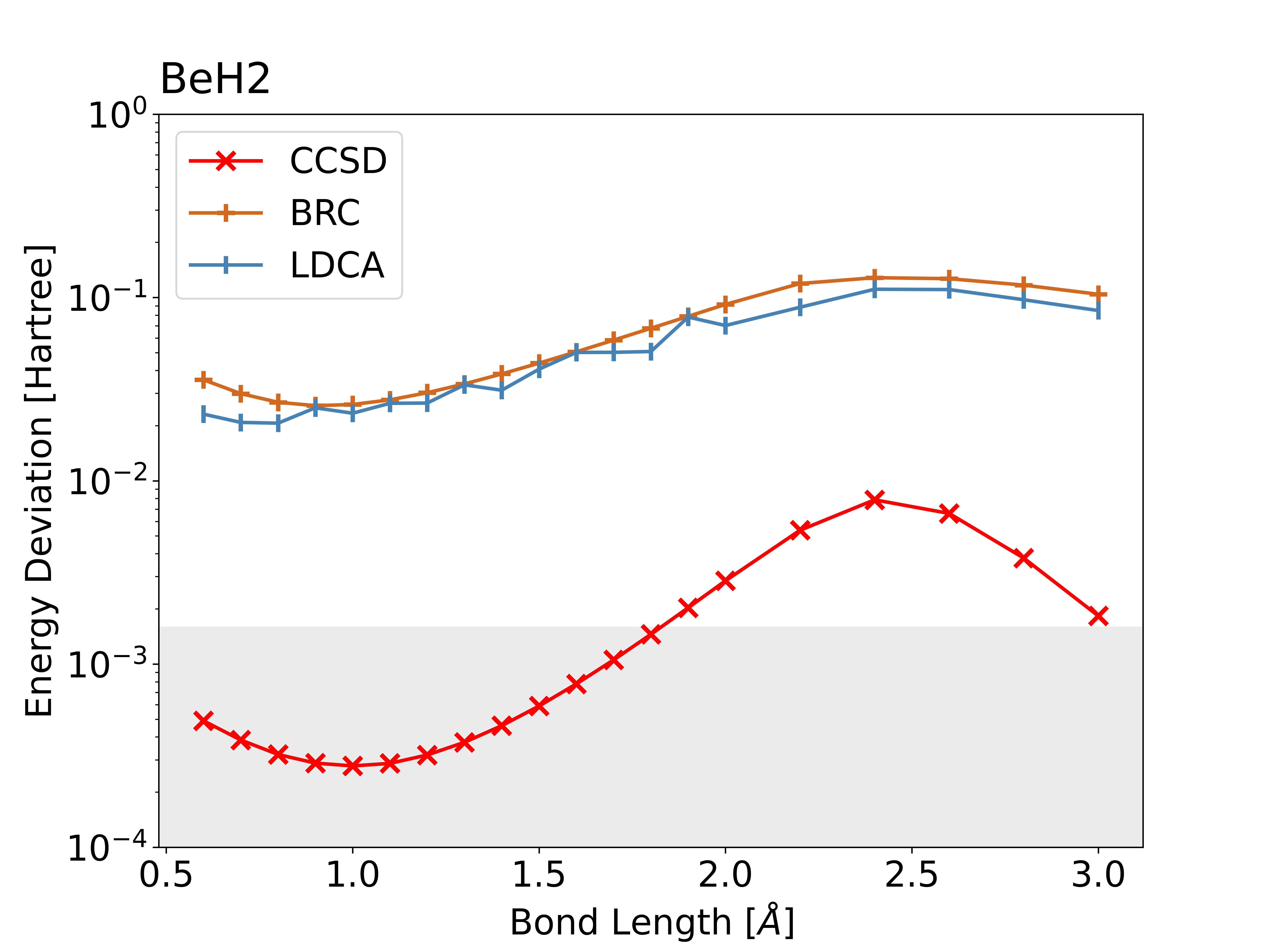}
		}
		\caption{
		The energy error of $\mol{H_4}, \mol{LiH}$ and $\mol{BeH_2}$ calculated by CCSD method, LDCA and BRC ansatzes compared to the full-CI method at different bond length. The results of full-CI method are settled as reference(0 energy). Some energy points calculated by CCSD are lower than full-CI energy, which can't be shown in the figure. The grey area denotes the chemical accuracy (-0.0016 $<$ energy error $<$ 0.0016). }
		\label{fig:energy3}
	\end{figure*}     
	
    The energy results calculated by BRC and LDCA are shown in Figure~\ref{fig:energy3}. The results show that both BRC and LDCA had remarkable energy errors and couldn't achieve chemical accuracy. This suggest BRC and LDCA may be needed to repeat more layers to reach needed accuracy.

  \subsection{Comparison of Runtime Until Convergence}\label{sec:runtime}
    The runtime of each case is discussed to indicate the hardness for different ansatzes simulated by the quantum simulator. It can reflect the cost of time when the ansatzes run on real quantum devices to some extent.
    
\begin{figure*} [htbp]
		\centering
		\noindent\makebox[\textwidth][c] {
		\hspace{9mm}
		\includegraphics[width=0.32\paperwidth]{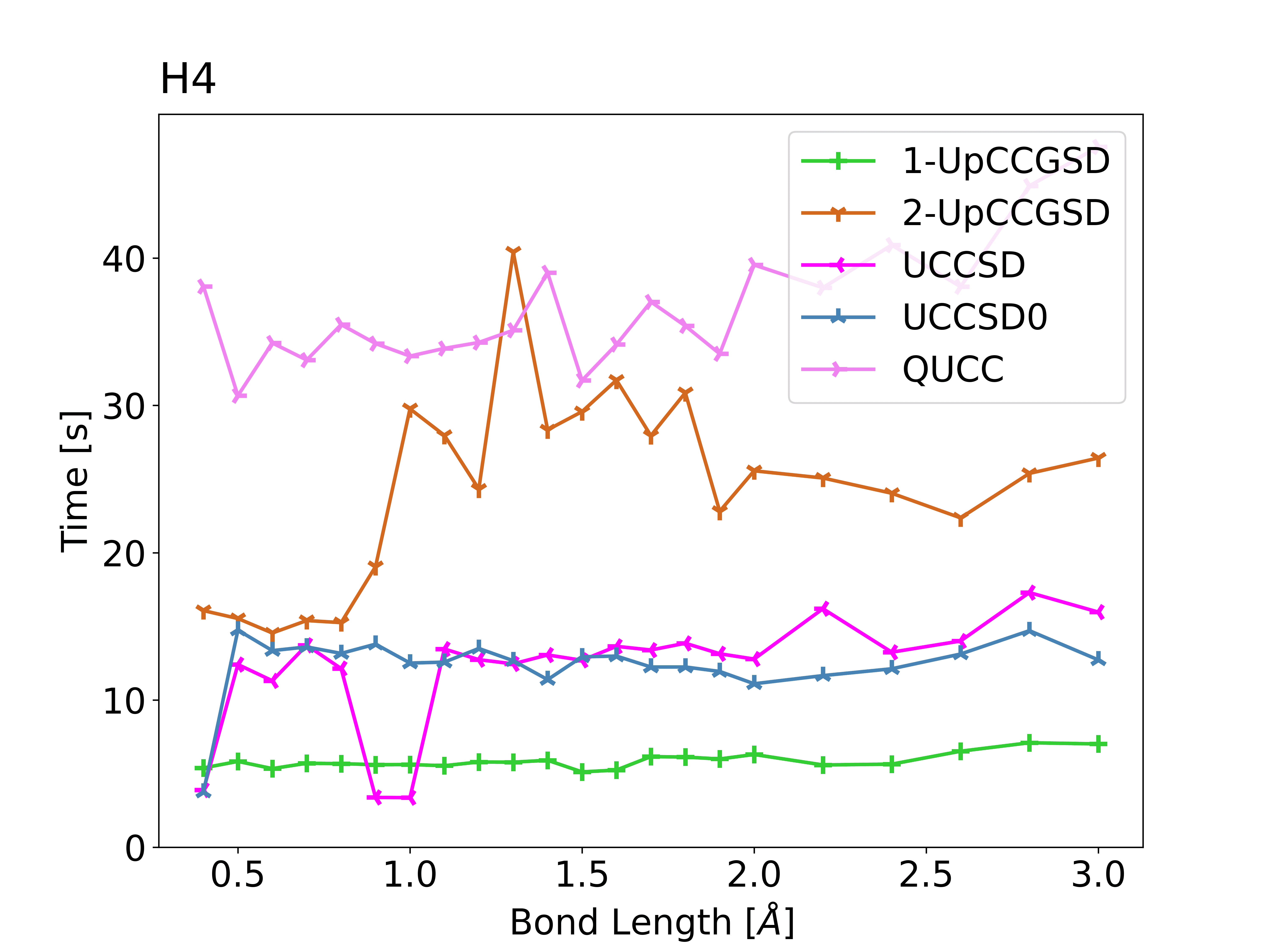}
		\hspace{-9mm}
		\includegraphics[width=0.32\paperwidth]{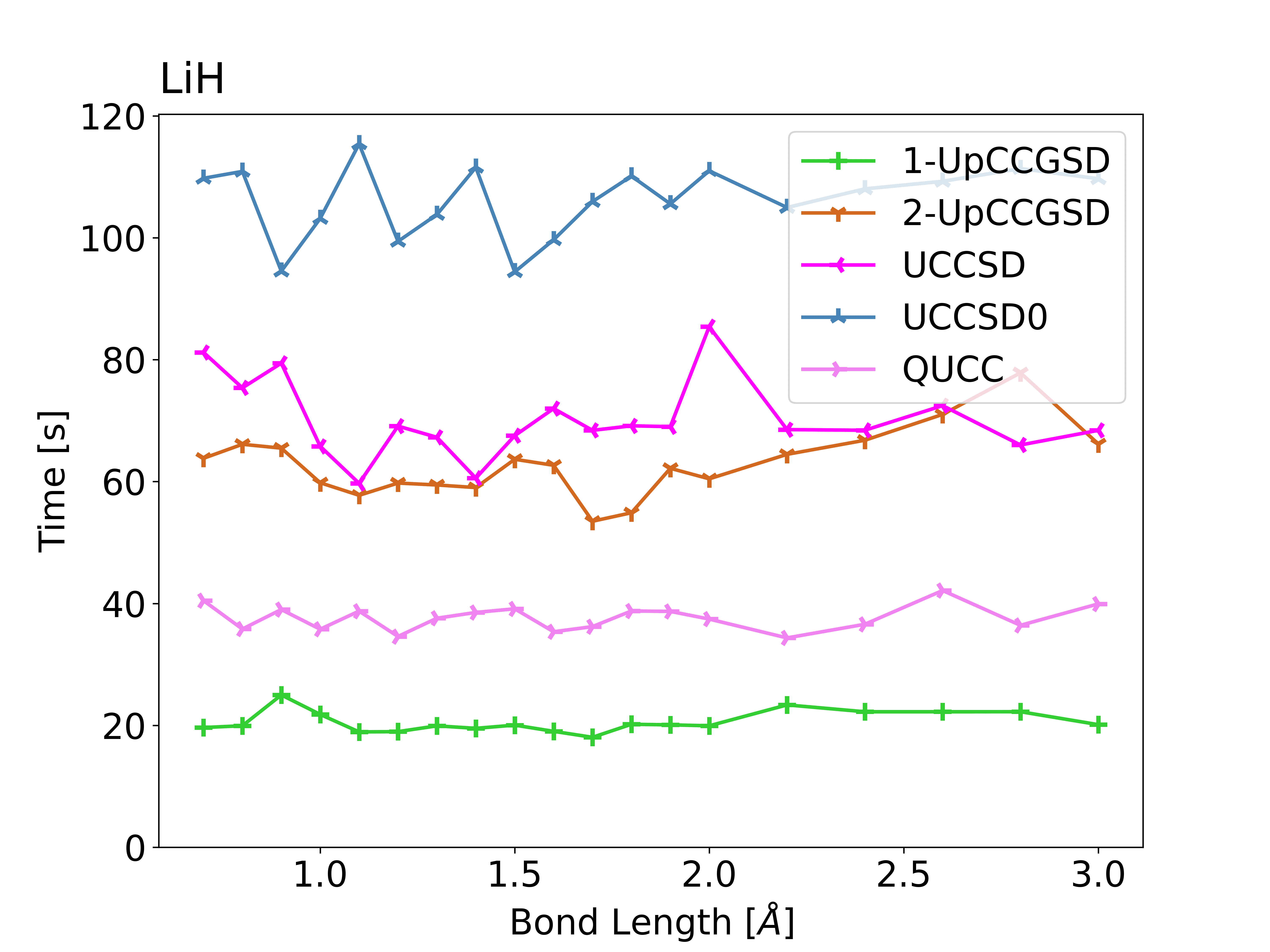}
		\hspace{-9mm}
		\includegraphics[width=0.32\paperwidth]{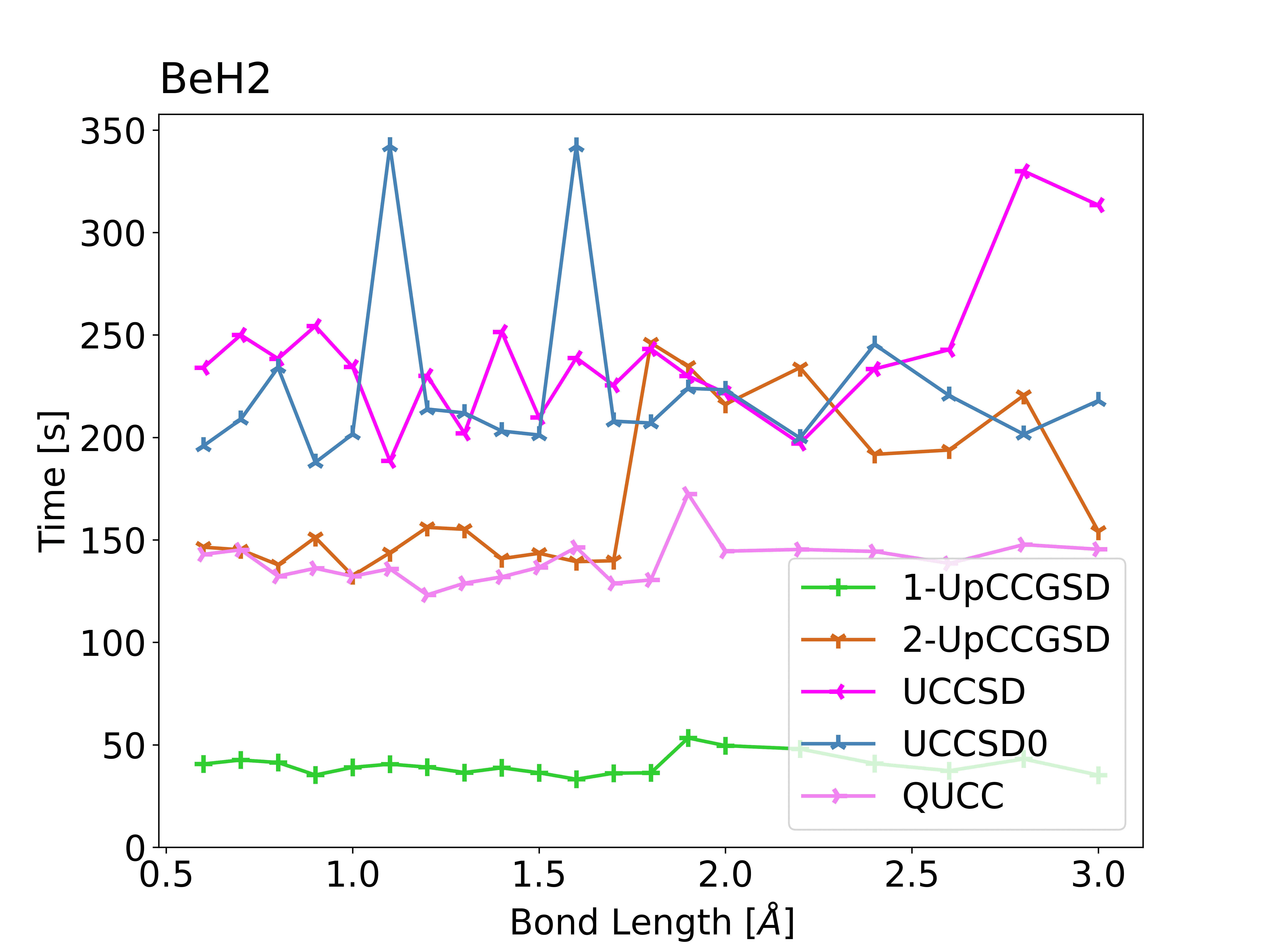}
		}
		\noindent\makebox[\textwidth][c] {
		\hspace{9mm}
		\includegraphics[width=0.32\paperwidth]{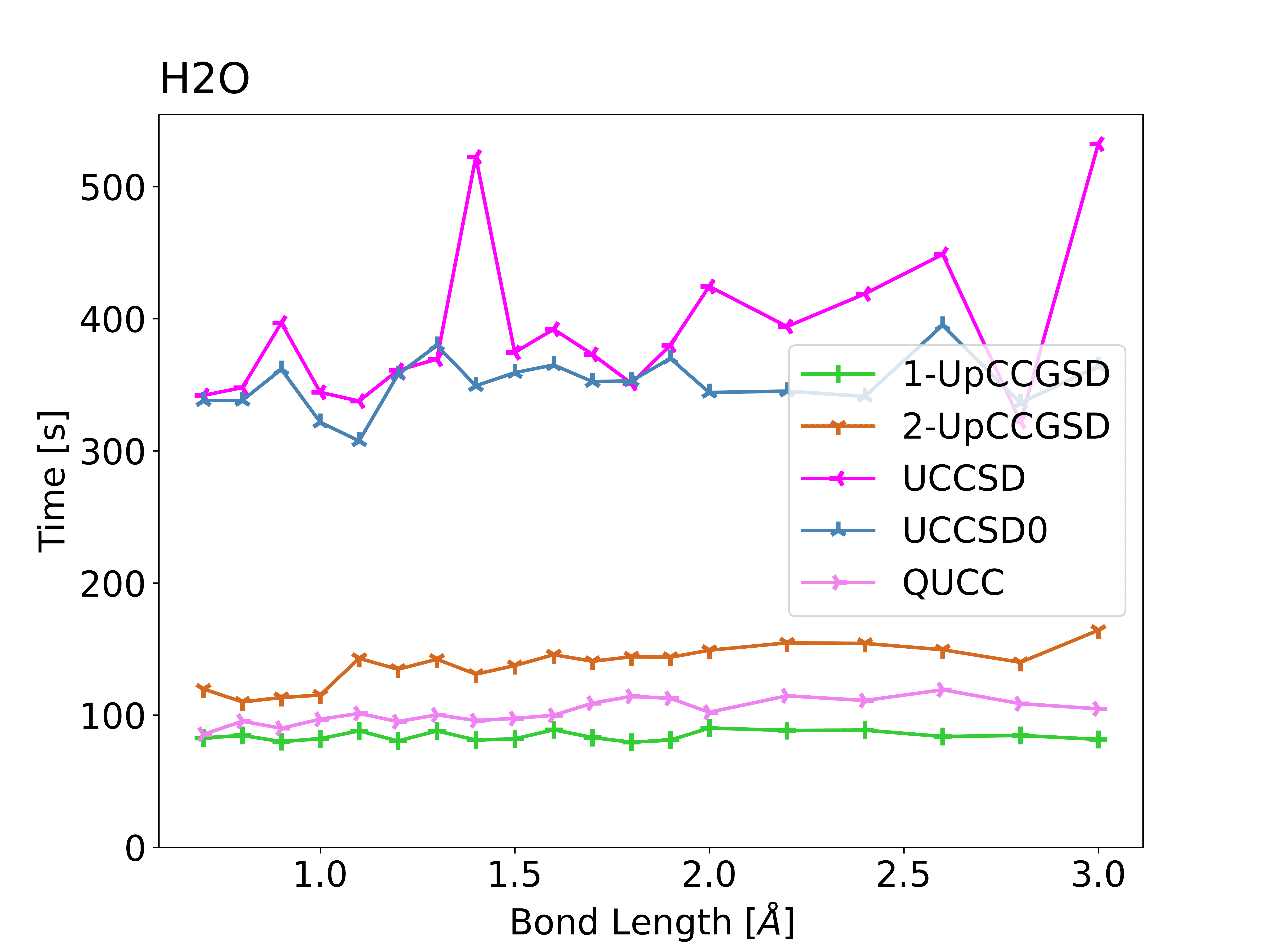}
		\hspace{-9mm}
		\includegraphics[width=0.32\paperwidth]{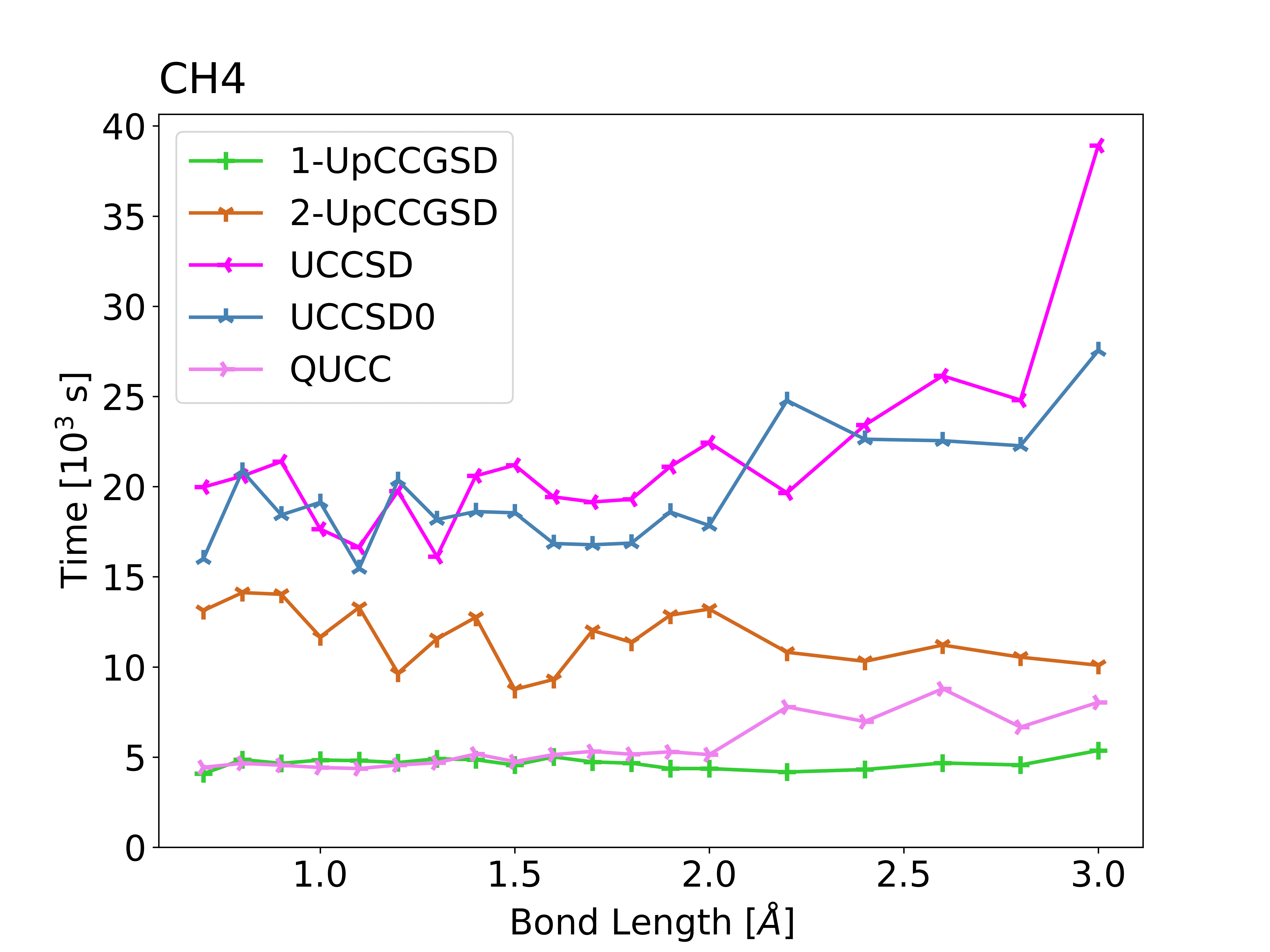}
		\hspace{-9mm}
		\includegraphics[width=0.32\paperwidth]{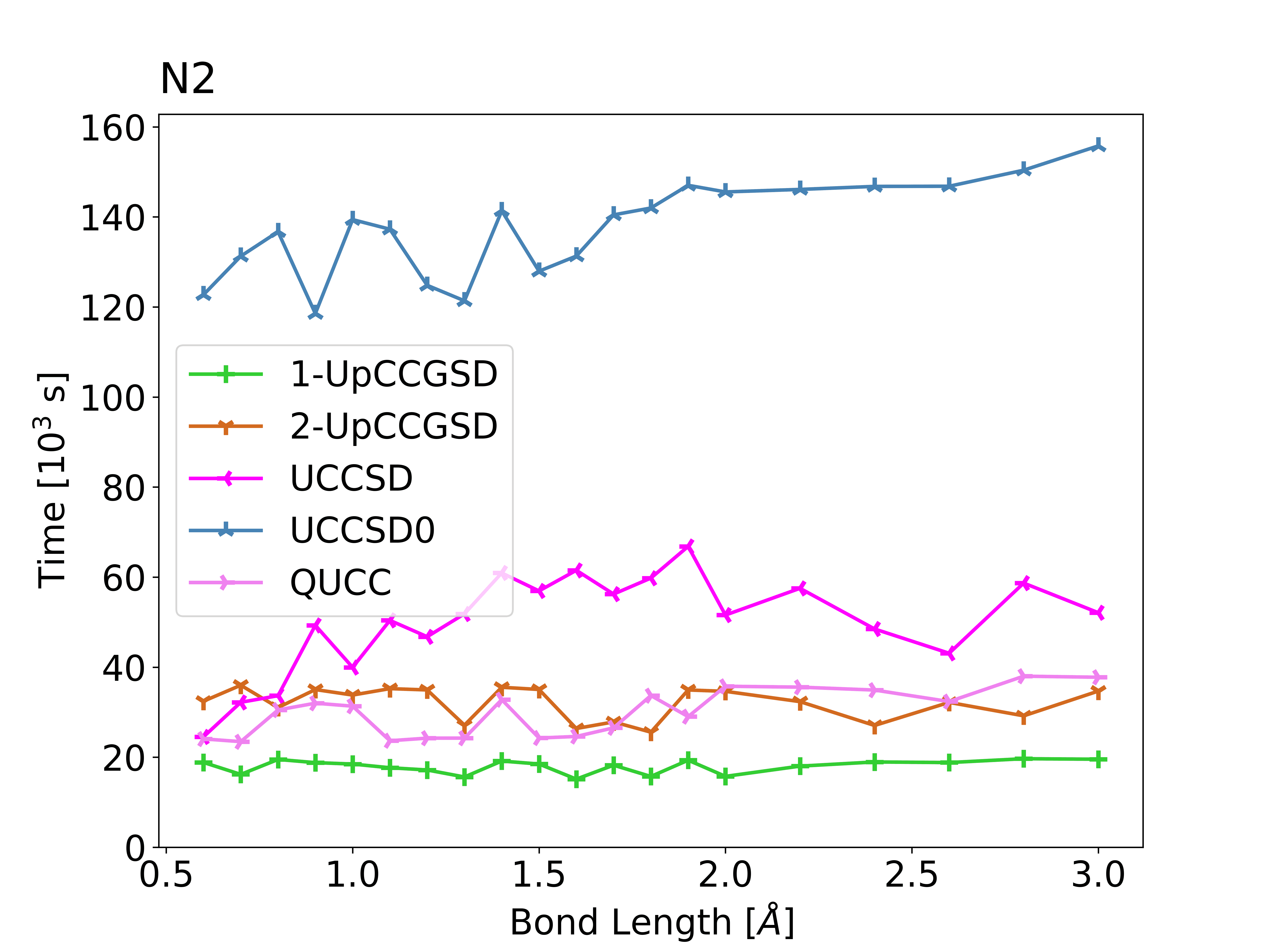}
		}
		\caption{The runtime of different fixed-circuit ansatzes to calculate $\mol{H_4}, \mol{LiH}, \mol{BeH_2}, \mol{H_2O}, \mol{CH_4}$ and $\mol{N_2}$ at different bond lengths until convergence.}
		\label{fig:time1}
	\end{figure*}

    The runtime of different fixed-circuit ansatzes is shown in Figure~\ref{fig:time1}. In general, except for the $\mol{H_4}$ molecule case, the ansatzes sorted in ascending order of runtime were 1-UpCCGSD, QUCC, 2-UpCCGSD, UCCSD, UCCSD0. From the results, the relationships between the runtime and bond lengths for each specific molecule are not obvious.
  
\begin{figure*} [htbp]
		\centering
		\noindent\makebox[\textwidth][c] {
		\hspace{9mm}
		\includegraphics[width=0.32\paperwidth]{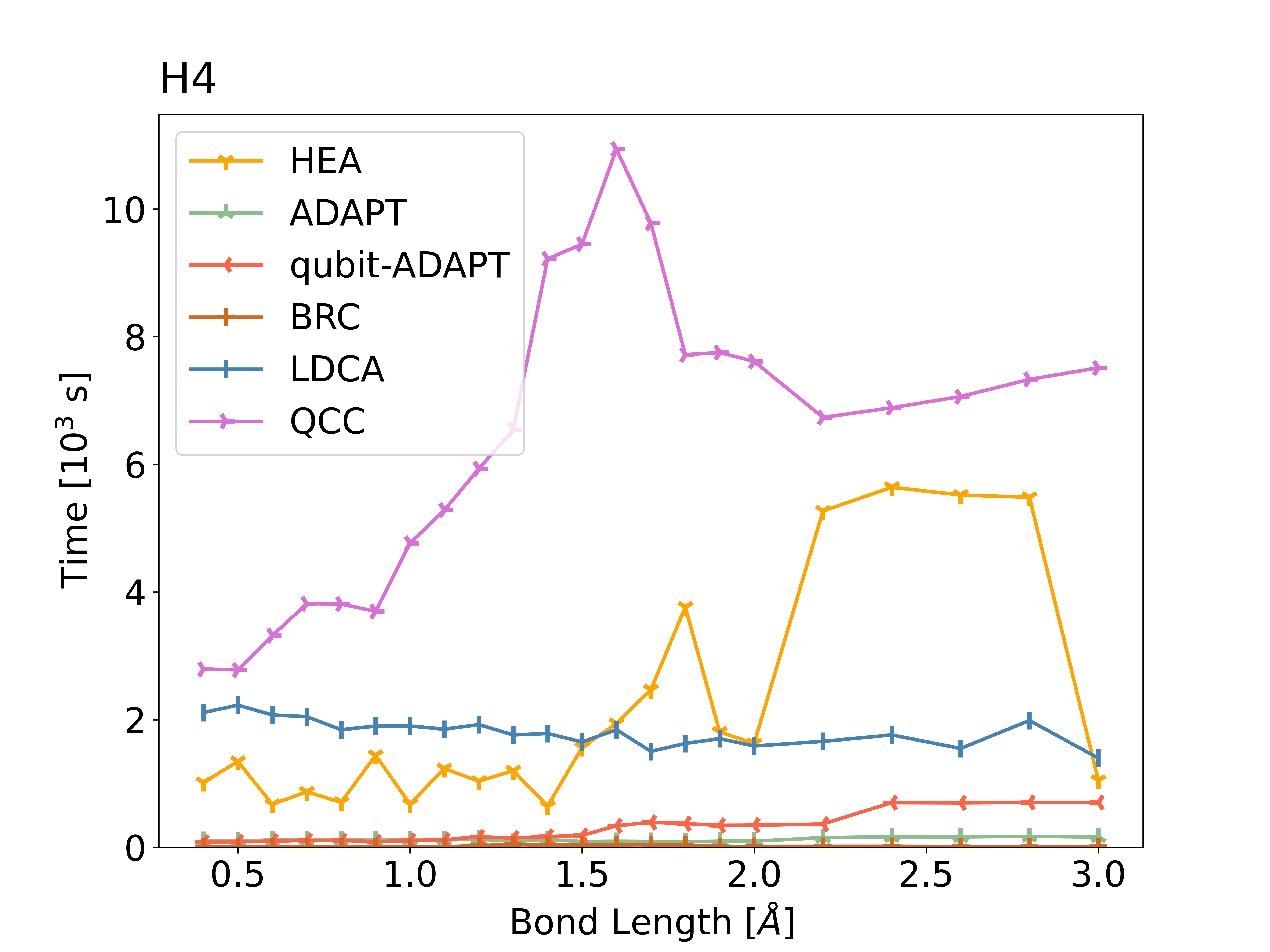}
		\hspace{-9mm}
		\includegraphics[width=0.32\paperwidth]{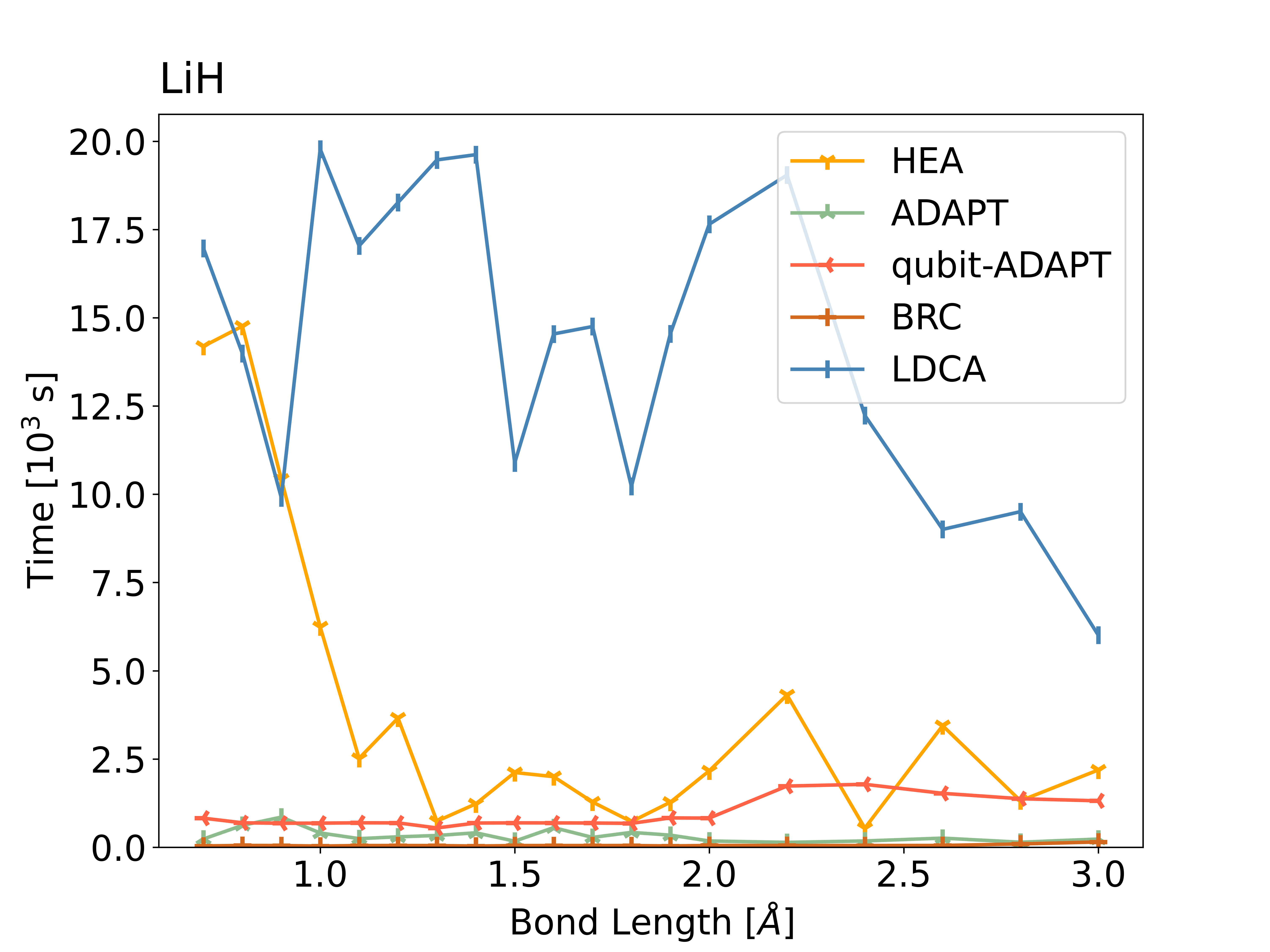}
		\hspace{-9mm}
		\includegraphics[width=0.32\paperwidth]{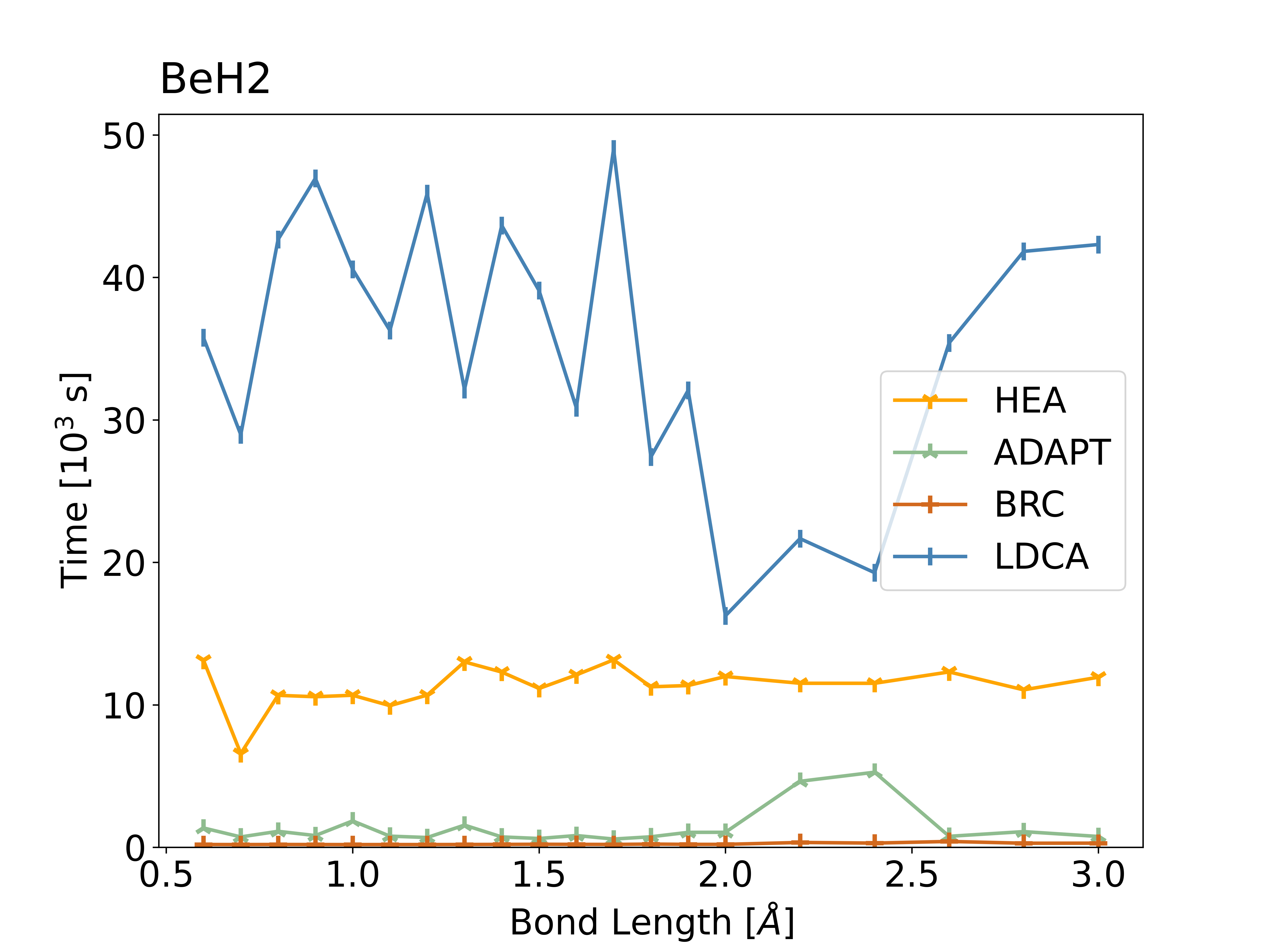}
		}
		\caption{The runtime of different changeable-circuit ansatzes to calculate $\mol{H_4}, \mol{LiH}, \mol{BeH_2}$ at different bond lengths until convergence.}
		\label{fig:time2}
	\end{figure*}

	The runtime of changeable-circuit ansatzes is shown in Figure~\ref{fig:time2}. Generally, the ansatzes sorted in ascending order of runtime was BRC, QCC, HEA, qubit-ADAPT, ADAPT, LDCA. This order was because the runtime of changeable-circuit ansatz was related to the size of the operator pool when screening the operators. 
	For some points of bond lengths in different molecules, respectively, the runtime happened to largely increased, of which the reason is still under exploration.
	
    To sum up, the changeable-circuit ansatzes usually cost more time in the classical simulator than the fixed-circuit ansatzes, as they cost extra computational resources to find the best circuit structure. By this reason, changeable-circuit ansatzes may not be practical in large systems.
	
  \subsection{Comparison of Number of Parameters}
    
    The number of variational parameters in different ansatzes was also explored. Usually, this indicates the difficulty of optimization and the circuit depth of ansatzes. 
    
\begin{figure*} [htbp]
		\centering
		\noindent\makebox[\textwidth][c] {
		\hspace{9mm}
		\includegraphics[width=0.32\paperwidth]{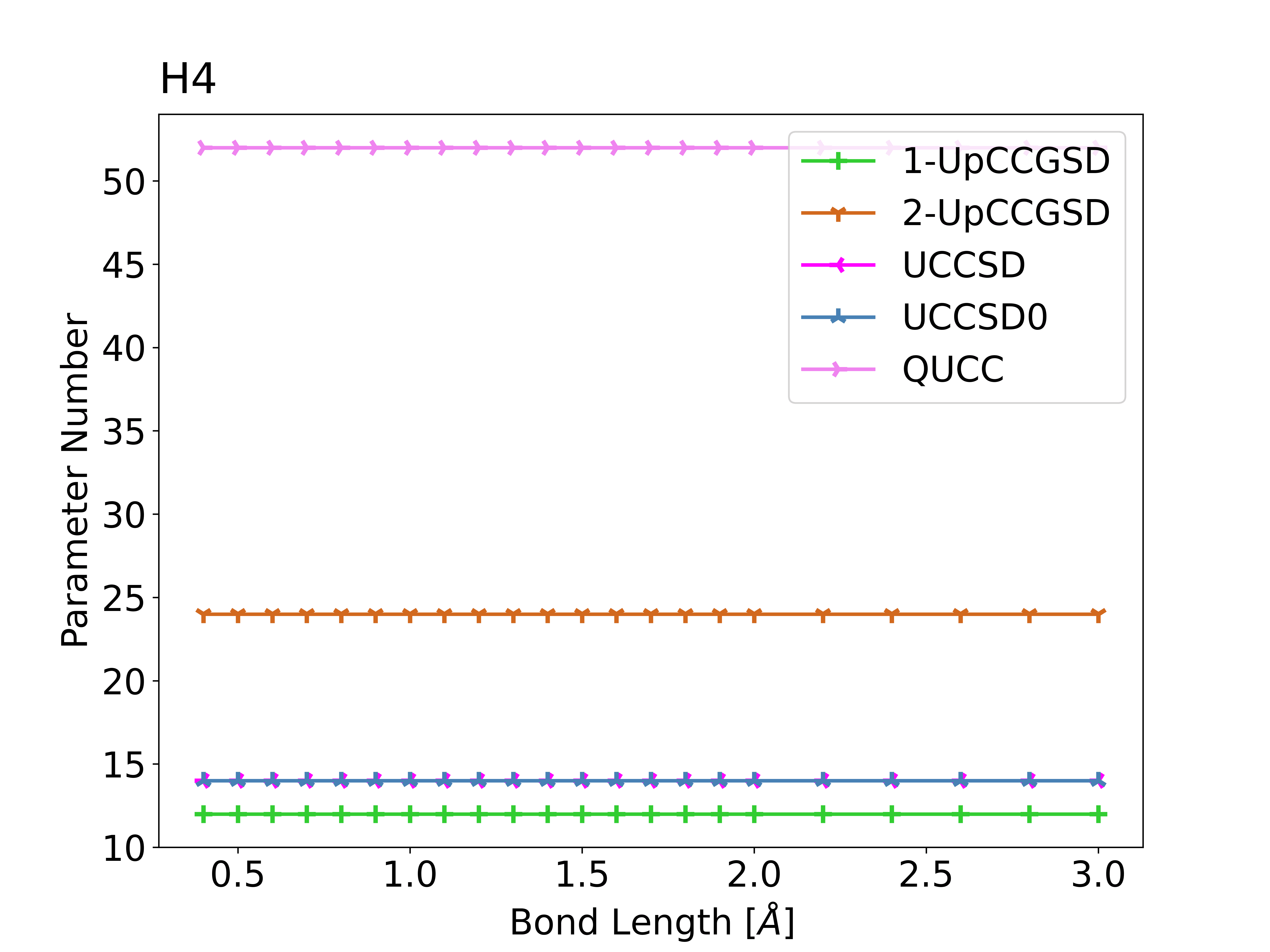}
		\hspace{-9mm}
		\includegraphics[width=0.32\paperwidth]{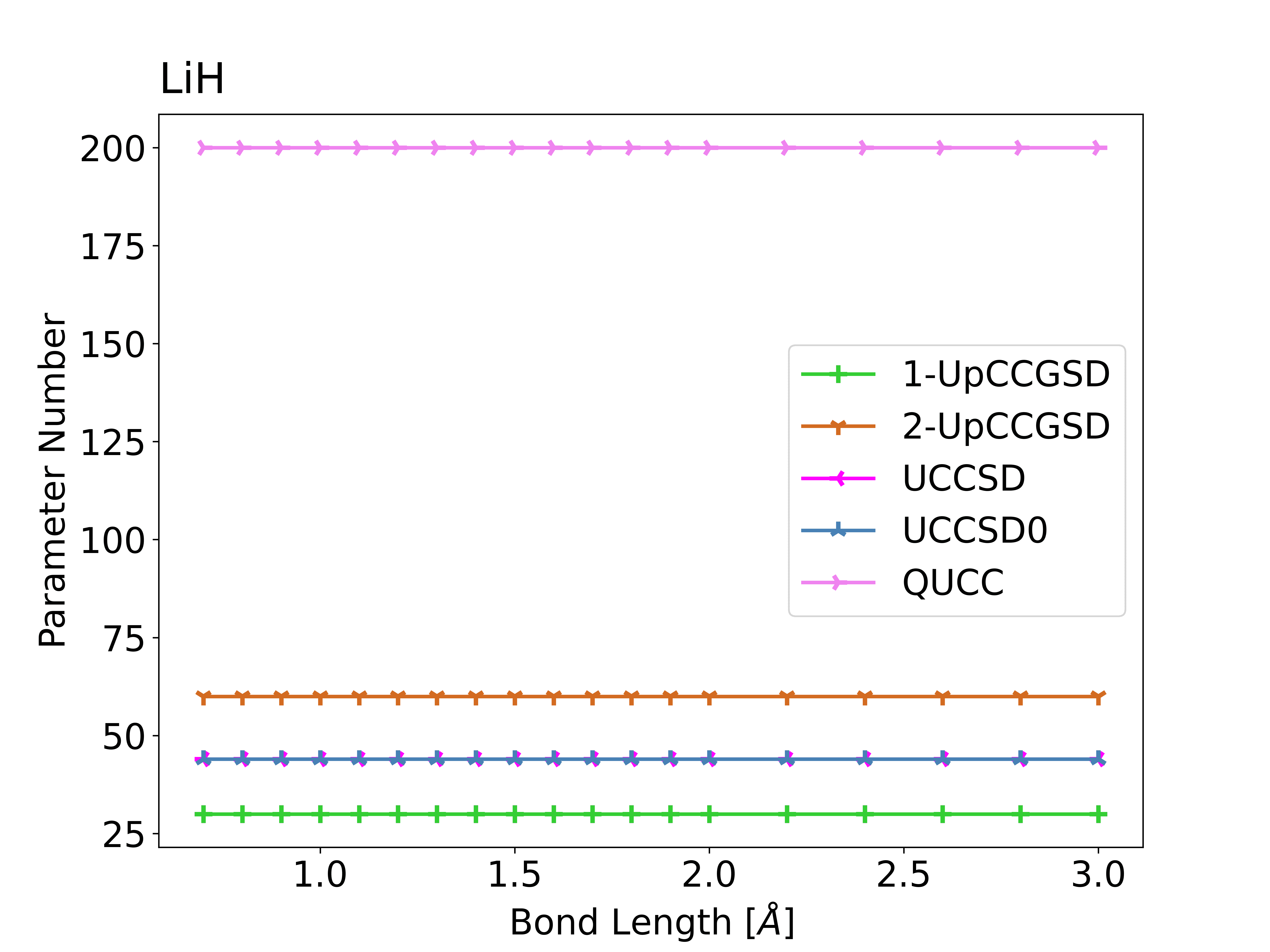}
		\hspace{-9mm}
		\includegraphics[width=0.32\paperwidth]{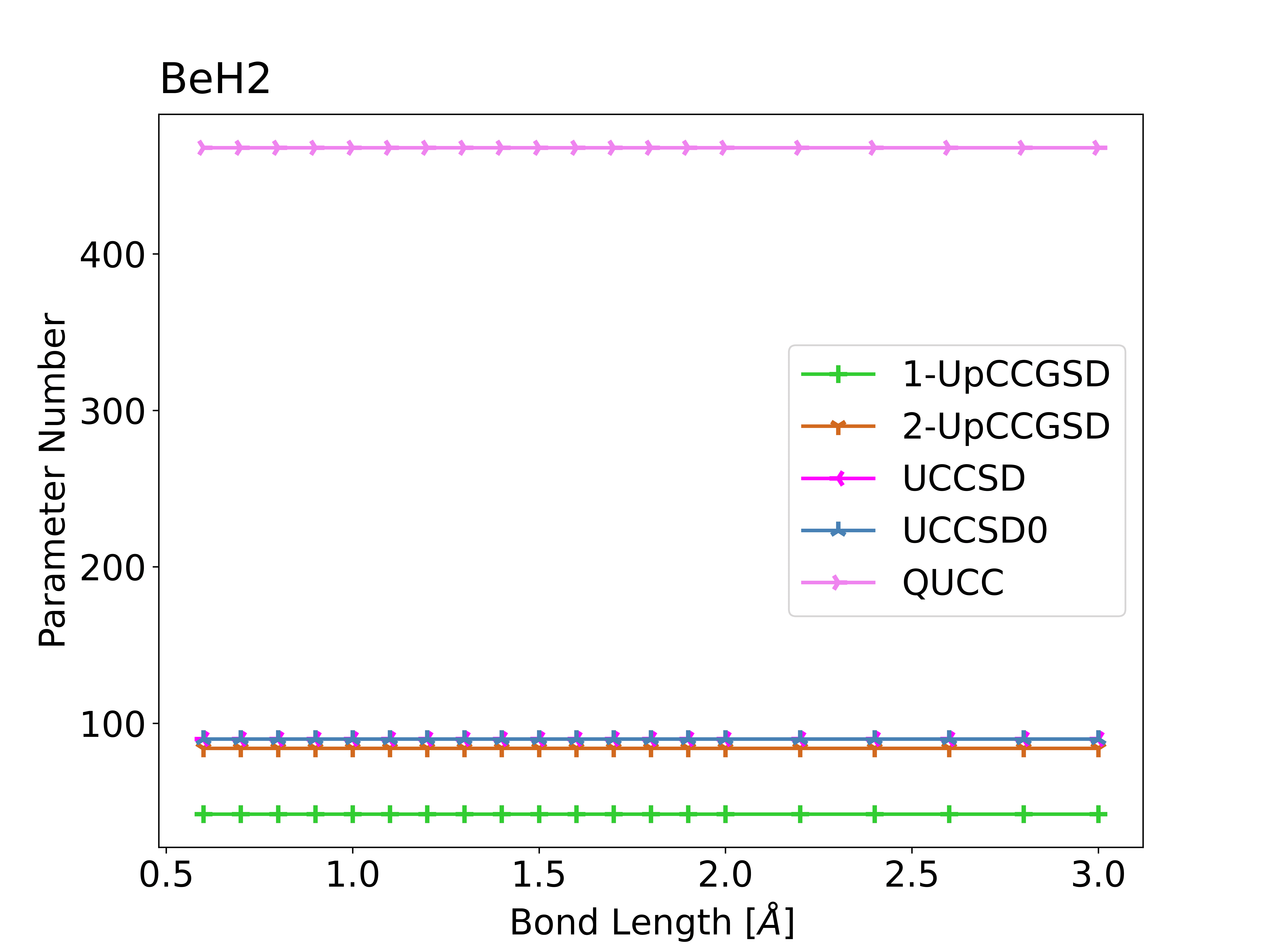}
		}
		\noindent\makebox[\textwidth][c] {
		\hspace{9mm}
		\includegraphics[width=0.32\paperwidth]{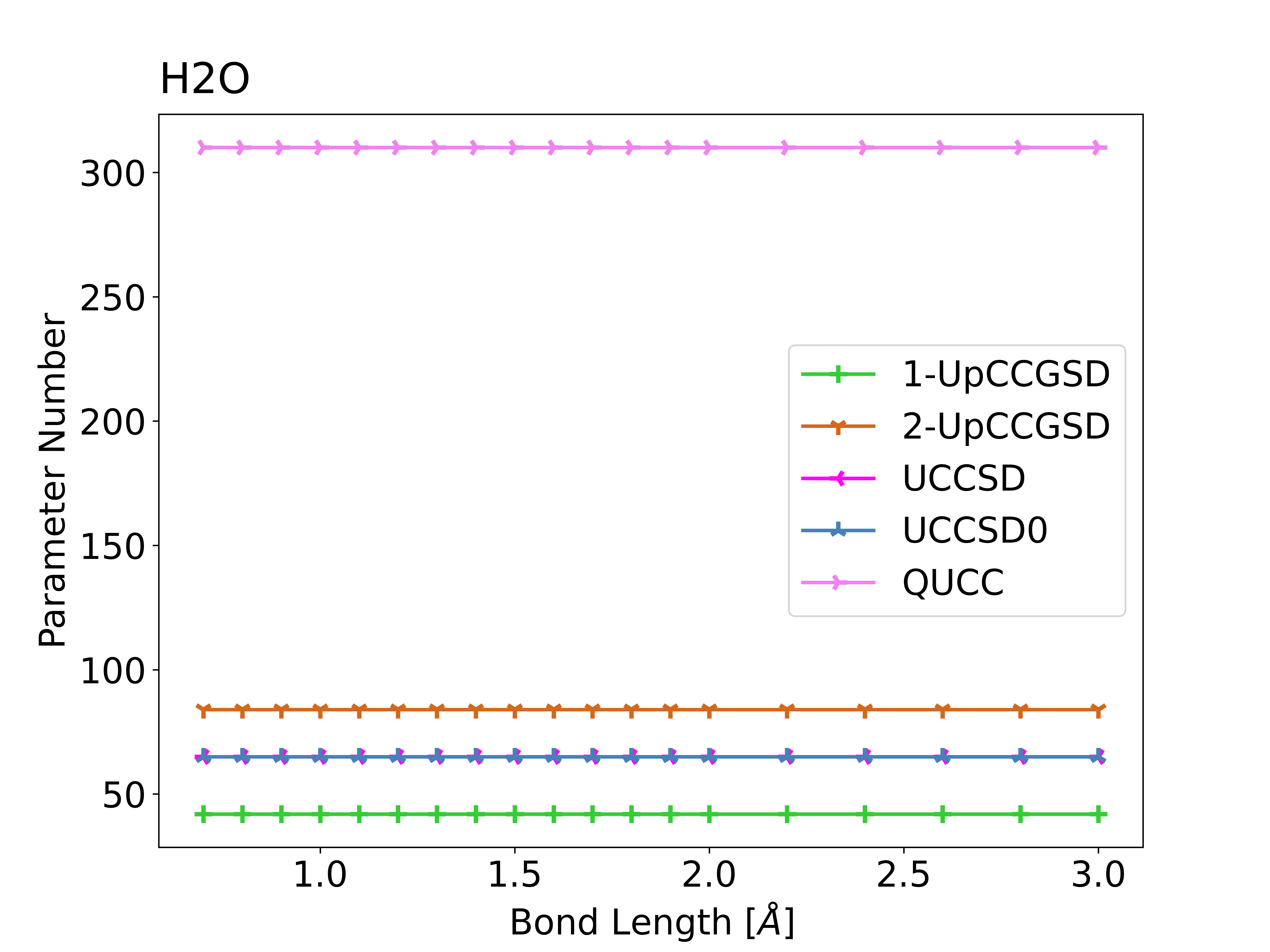}
		\hspace{-9mm}
		\includegraphics[width=0.32\paperwidth]{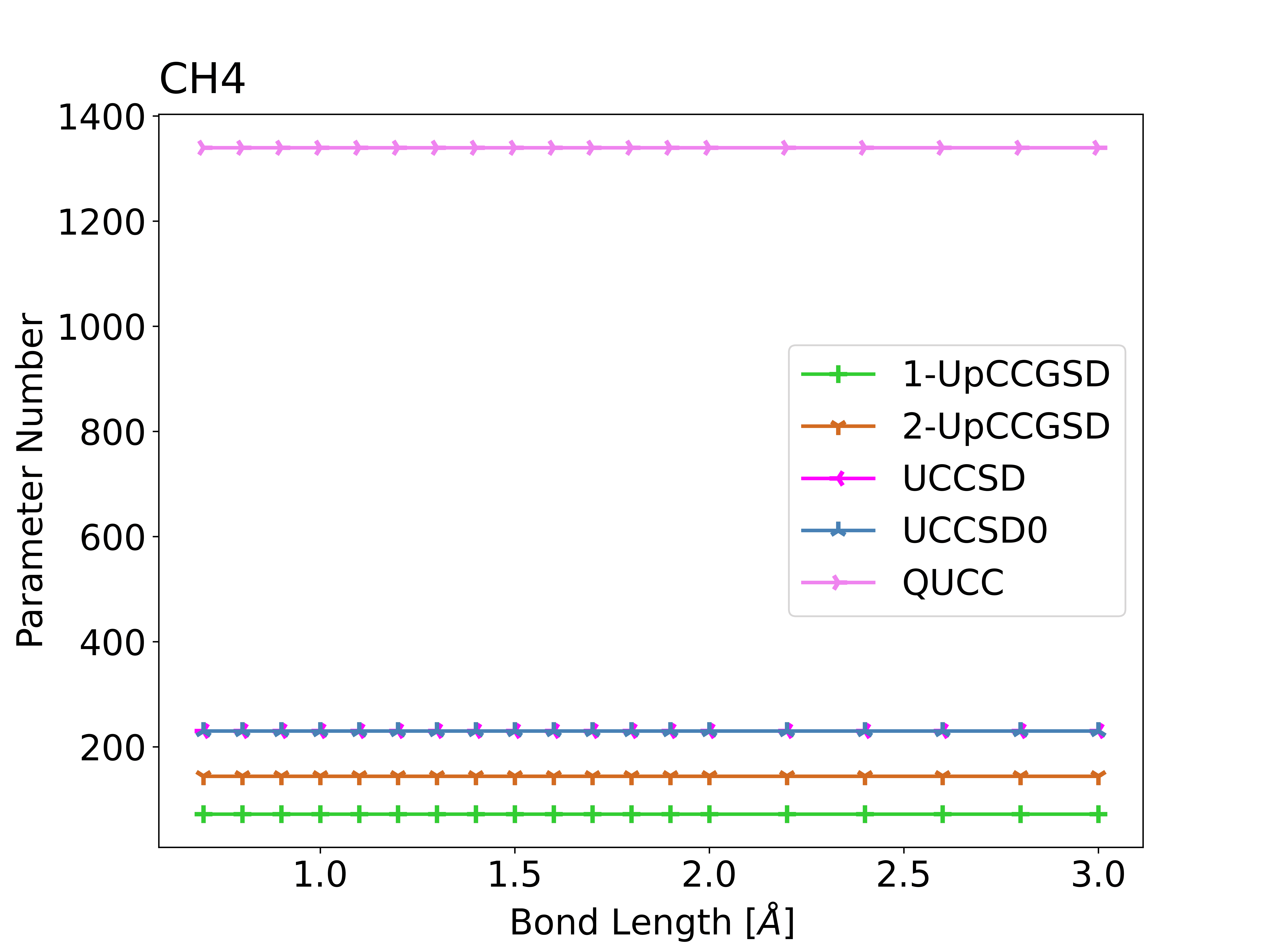}
		\hspace{-9mm}
		\includegraphics[width=0.32\paperwidth]{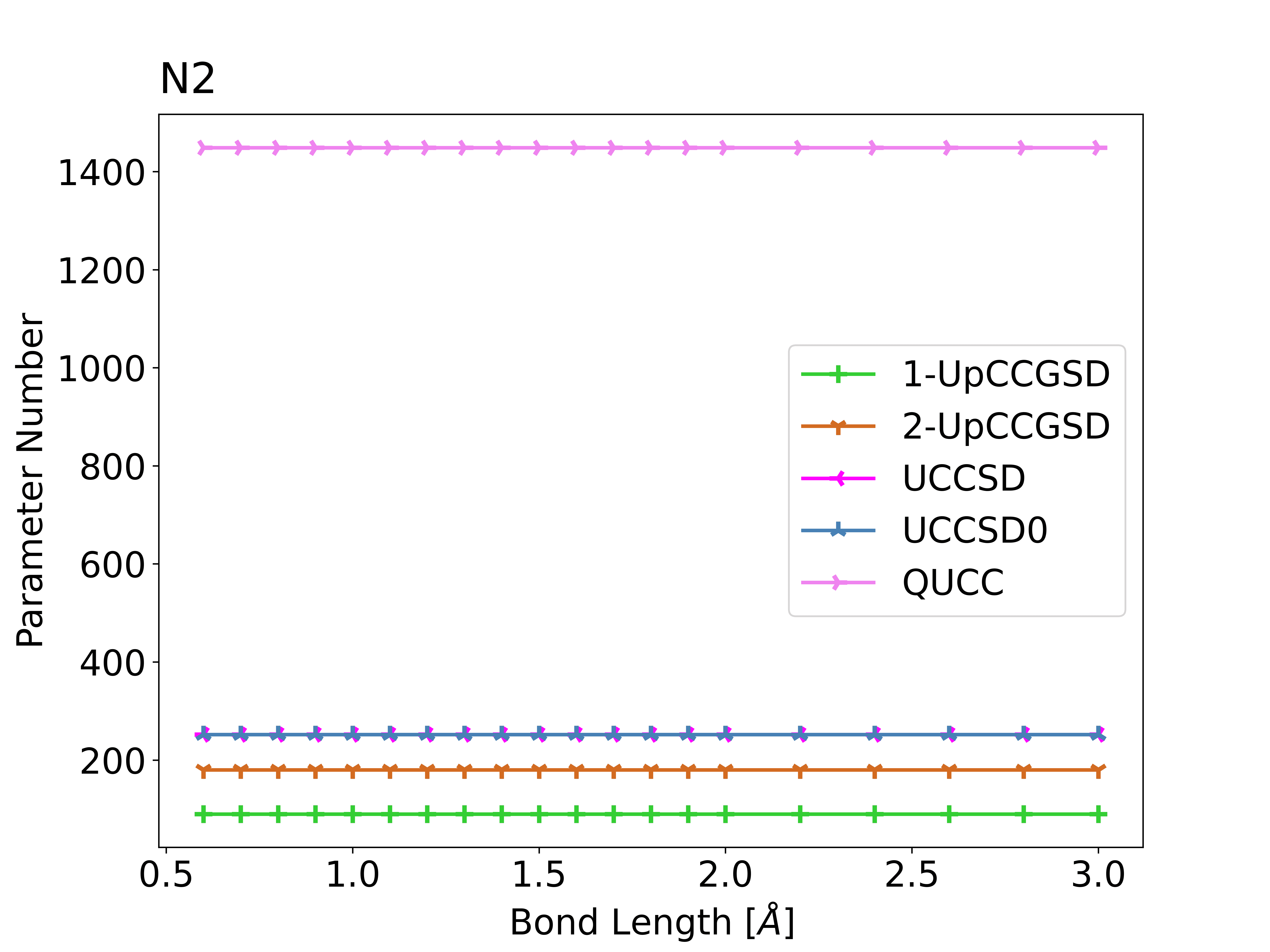}
		}
		\caption{The number of parameters used by different fixed-circuit ansatzes to calculate $\mol{H_4}, \mol{LiH}, \mol{BeH_2}, \mol{H_2O}, \mol{CH_4}$ and $\mol{N_2}$. The number will not change with the bond length. The number of parameters of the UCCSD ansatz is always the same with the UCCSD0 ansatz.}
		\label{fig:parameter1}
	\end{figure*}

\begin{figure*} [htbp]
		\centering
		\noindent\makebox[\textwidth][c] {
		\hspace{9mm}
		\includegraphics[width=0.32\paperwidth]{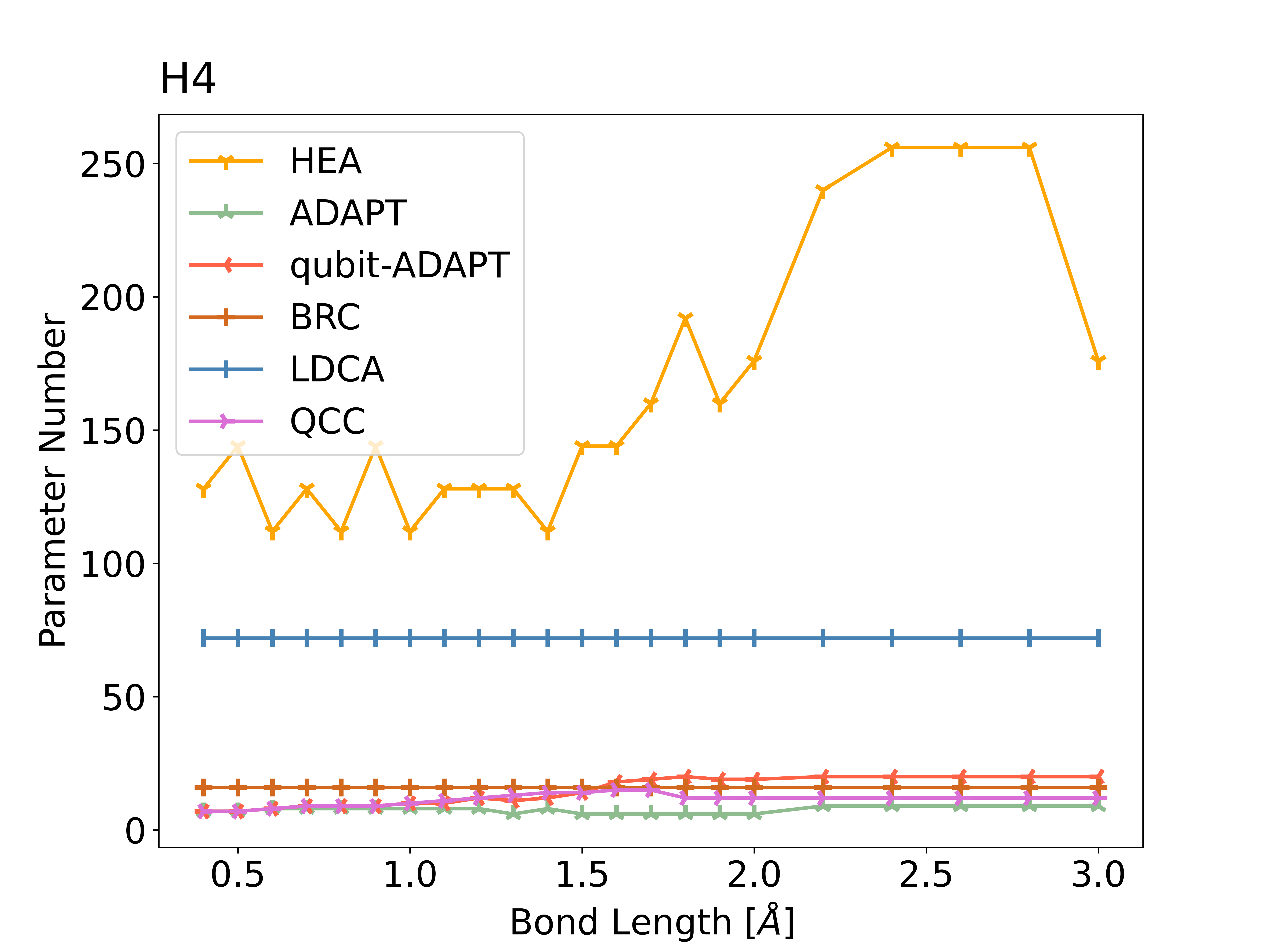}
		\hspace{-9mm}
		\includegraphics[width=0.32\paperwidth]{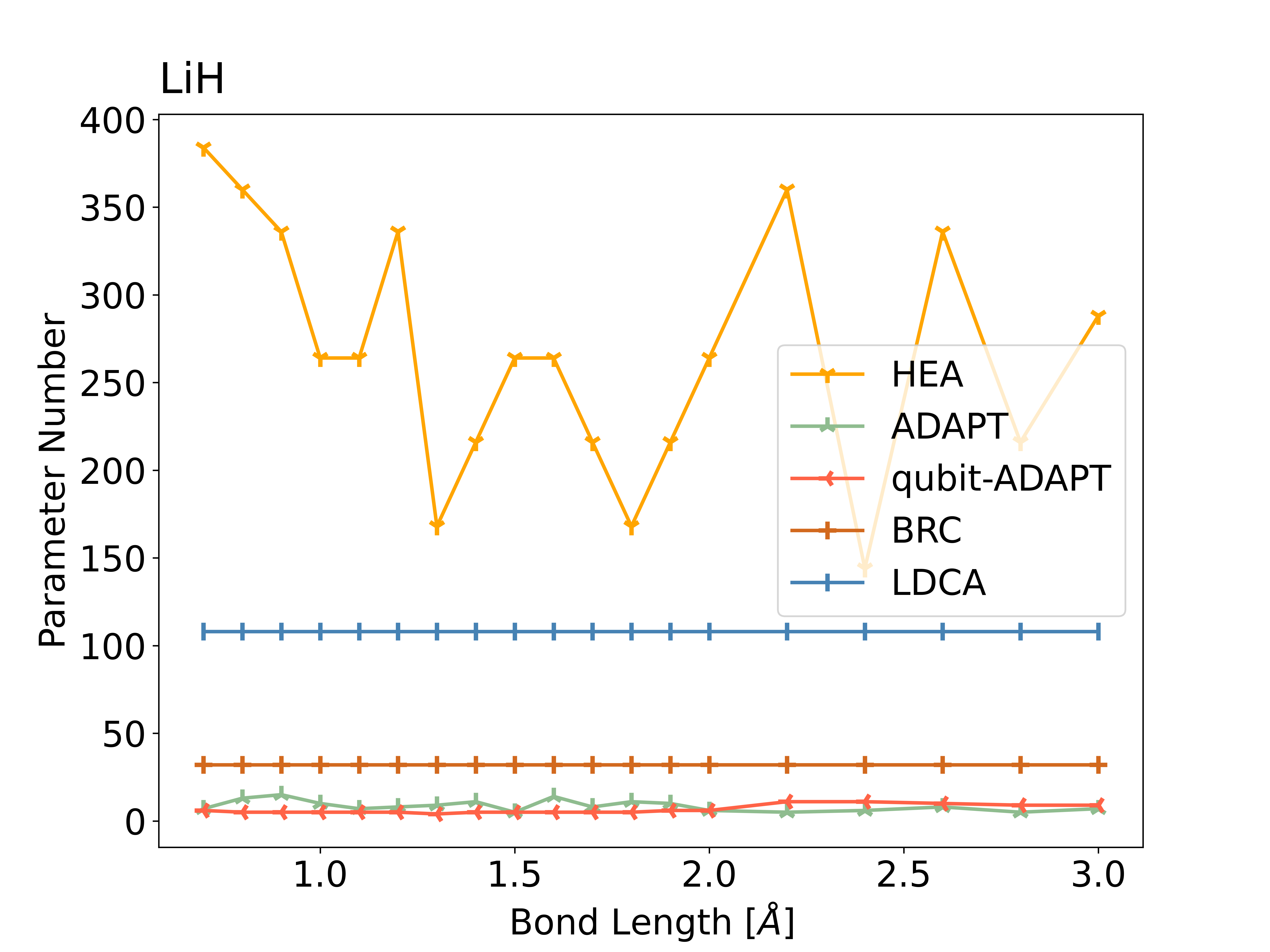}
		\hspace{-9mm}
		\includegraphics[width=0.32\paperwidth]{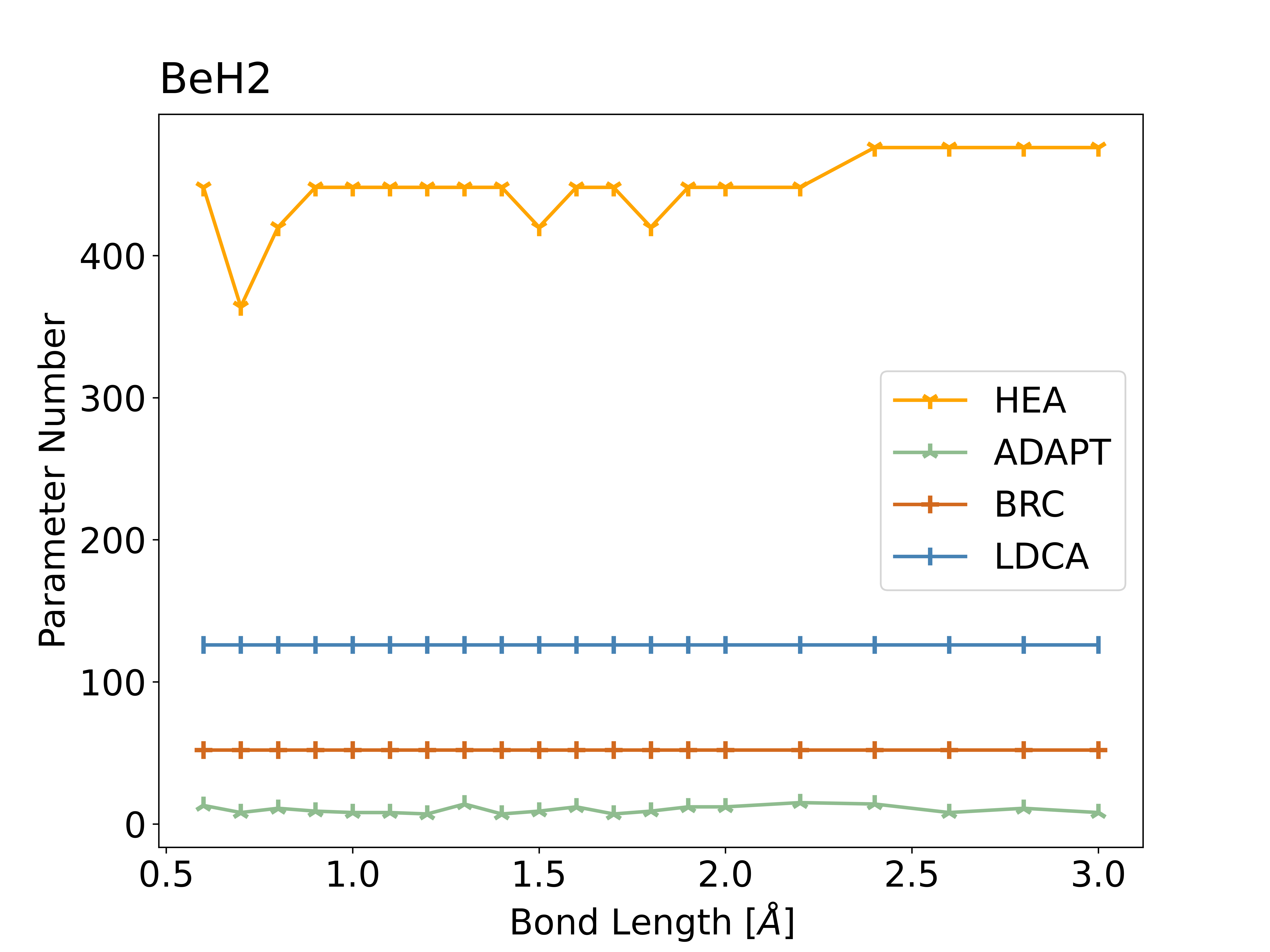}
		}
		\caption{The number of parameters used by different changeable-circuit ansatzes to calculate $\mol{H_4}, \mol{LiH}, \mol{BeH_2}$.}
		\label{fig:parameter2}
	\end{figure*}

    The number of variational parameters of fixed-circuit ansatzes are shown in Figure~\ref{fig:parameter1}. Generally, sorted in ascending order of the number of parameters they contained, the rank of these ansatzes was 1-UpCCGSD, UCCSD(UCCSD0), 2-UpCCGSD, QUCC. However, when the size of the system grew, the 2-UpCCGSD method requires fewer parameters than UCCSD, suggested that one may try to use higher-order k-UpCCGSD ansatz for larger systems. 
	The number of parameters used by different changeable-circuit ansatzes is shown in Figure~\ref{fig:parameter2}. In changeable-circuit ansatzes, the number of parameters one ansatz used is associated with the iteration steps of each ansatz, and also depends on the convergence conditions.
	
    In a word, adaptive ansatzes, inclucing ADAPT and qubit-ADAPT, require less variational parameters and smaller circuit depths than LDCA, HEA and fixed-circuit ansatzes, which means they have relatively shallower circuit depth. However, as mentioned in Section~\ref{sec:runtime}, the optimizations of adaptive ansatzes may not be easier than fixed-circuit ansatzes. Considering the number of parameters used by these ansatzes, it is recommended to use UCCSD0 or k-UpCCGSD ansatzes for large systems.

\section{Discussion} \label{sec:discussion}
  Using the MindQuantum software package~\cite{MindQuantum}, we benchmarked the performance of different VQE ansatzes, including energy error, runtime until convergence, and number of parameters. Due to limited space, we presented only partial results, readers can readily reproduce the results and perform new testing by applying our toolkit. The codes are publicly available at Gitee~\cite{MindQuantum}. 
  
  By analyzing the results, we conclude that most published ansatzes can reach chemical accuracy at the equilibrium points of the tested molecules. However, all the ansatzes failed to have reliable performance at stretched bond lengths. UCCSD0 and QUCC ansatzes can usually be used to obtain more accurate energy compared with other ansatzes, and UCCSD0 is more accurate than QUCC for larger systems. The ADAPT ansatz has relatively better performance in smaller molecular systems (below 14 qubits). It usually needs fewer variational parameters, but it is not scalable for larger molecules. Moreover, we can also found that changeable-circuit ansatzes usually cost much more time in classical simulators than fixed-circuit ansatzes. Our benchmarking results indicating a requirement of ansatz improvement at stretched bond length cases.
  
\begin{figure*} [htbp]
		\centering
		\includegraphics[width=0.5\paperwidth]{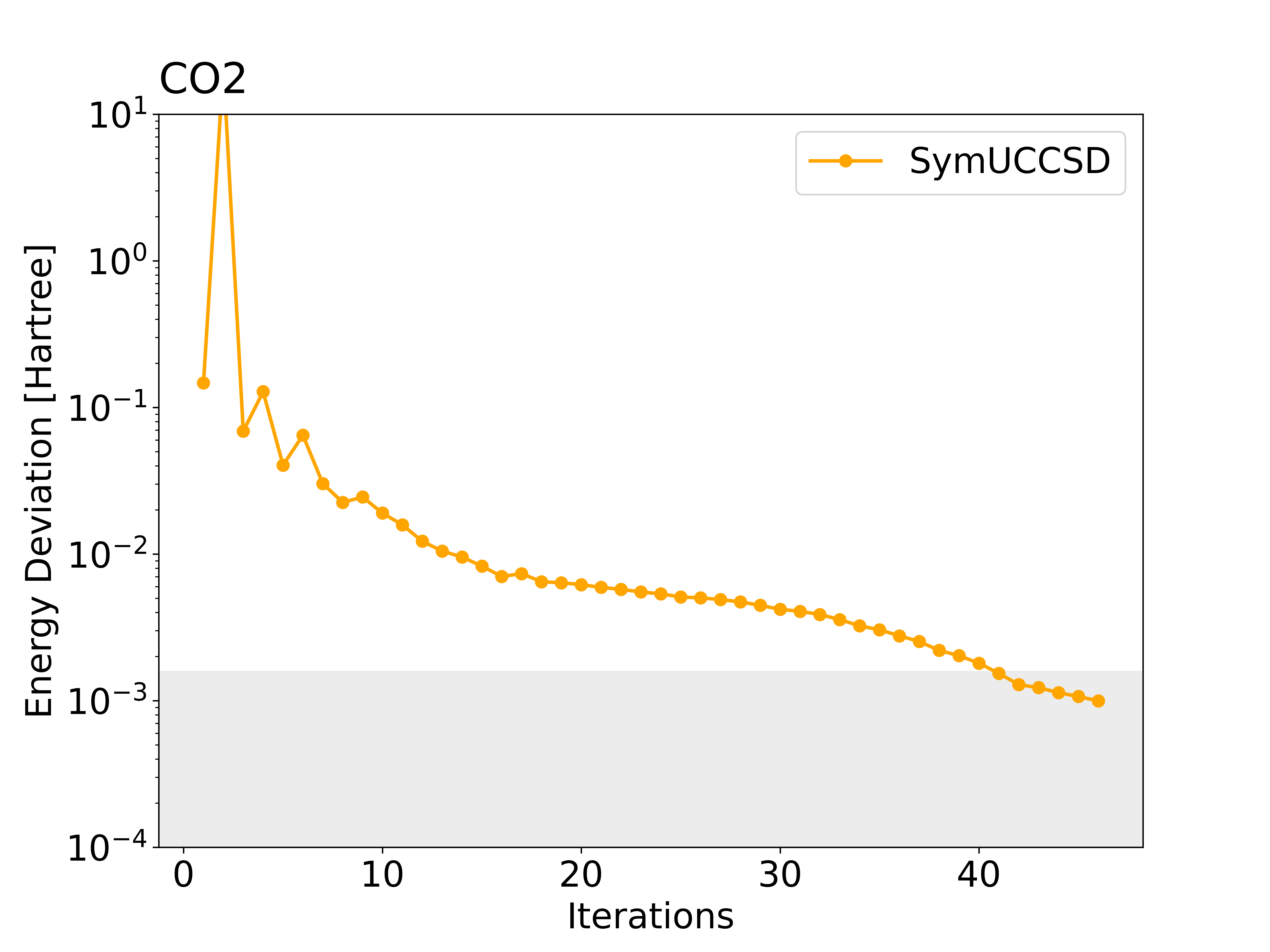}
		\caption{The energy convergence of $\mol{CO_2}$ molecule simulated by SymUCCSD ansatz\cite{Cao2021}. The full-CI energy is set to be zero, and the gray area denotes the chemical accuracy. The calculated energy reached chemical accuracy at the 41-th iteration.}
		\label{fig:co2}
	\end{figure*}

  Apart from the above results, we also perform large-scale VQE simulations using the MindQuantum package. In the current version (v0.6) of MindQuantum, the same 28-qubit simulation of the $\mol{C_2H_4}$ molecule~\cite{Cao2021} can be executed in much less time: each iteration can be reduced from 960 minutes to 2.5 minutes. To take a step further, we also deploy MindQuantum to benchmark up to 30 qubits $\mol{CO_2}$, which is by far the largest scale simulation in the literature, as far as we are aware of. The energy iteration results are shown in  Figure~\ref{fig:co2}.
  
  There are still many other cases that are meaningful to be benchmarked. Besides, there are also lots of ansatzes being proposed constantly. We will benchmark more ansatzes and update the results continuously. We also welcome more researchers to join this project. We hope these results can help the development of practical VQE applications.

\section*{Acknowledgement}
  \textbf{Funding:} This work was supported by the Key-Area Research and Development Program of Guang-Dong Province (Grant No. 2018B030326001), the National Natural Science Foundation of China (Nos. U1801661 and 12004162), the Guangdong Provincial Key Laboratory (Grant No. 2019B121203002), the Science, Technology and Innovation Commission of Shenzhen Municipality (JCYJ20170412152620376, KYTDPT20181011104202253), and the Natural Science Foundation of Guangdong Province (Grant No.2017B030308003).
  \textbf{Competing Interests:} The authors declare no competing interests.
  \textbf{Author list:} Jiaqi Hu$^*$, Junning Li, Yanling Lin, Hanlin Long, Xu-Sheng Xu, Zhaofeng Su, Wengang Zhang, Yikang Zhu, Man-Hong Yung$^*$.
  \textbf{Author Contributions:} J.H, W.Z, Y.L, Y.Z wrote the benchmarking code. H.H, J.H, H.L assisted with data collection. J.H, W.Z, H.L, Y.L, Y.Z wrote the paper. M-H.Y, Z.S revised the paper.
  \textbf{Data and materials availability:} The code used for the project and all the data can be found in MindQuantum pacakge, located at \url{https://gitee.com/mindspore/mindquantum}.

\clearpage
{\small

\bibliography{main}
\bibliographystyle{unsrt}
}

\clearpage

\appendix
\pagenumbering{roman}

  \section{}
  \begin{table*}[h!]
    \centering
    \caption{A summary of symbols.}
    \label{glossary}
    \begin{tabular}[H]{ p{3cm} p{12cm} }
    \hline
    $\hat{H}$ & Molecular Hamiltonian\\
    $N_e$ & The number of electrons in the molecule of interest\\
    $N_n$ & The number of nuclei in the molecule of interest\\
    $Z$ & Charges of a nucleus\\
    $\mathbf{R}$ & Three-dimensional coordinates of a nucleus\\
    $\mathbf{r}$ & Three-dimensional coordinates of a electron\\
    $\{h_{pq}\}$ & one-electron integrals of the Hamiltonian in the second quantized form\\
    $\{h_{pqrs}\}$ & two-electron integrals of the Hamiltonian in the second quantized form\\
    $\hat{P}$ & A Pauli string\\
    $\hat{\sigma}^x$ &  The Pauli x operator\\
    $\hat{\sigma}^y$ &  The Pauli y operator\\
    $\hat{\sigma}^z$ &  The Pauli z operator\\
    $\hat{\sigma}^I$ &  The Pauli identity operator\\
    $\ket{\Psi(\boldsymbol{\theta})}$ & A quantum state with tunable parameters $\boldsymbol{\theta}$\\
    $\ket{\Psi_0}$ & An initial state\\
    $E_{g}$ & The ground state energy of the problem Hamiltonian\\
    $\hat{a}^\dagger$ & Fermionic creation operator\\
    $\hat{a}$ & Fermionic annihilation operator\\
    $\hat{T}(\boldsymbol{\theta})$ & The coupled-cluster excitation operators\\
    $\hat{t}_i^a$ & Single excitation operator that excites one electron from occupied spin orbital $i$ to virtual spin orbital $a$\\
    $\hat{t}_{ij}^{ab}$ & Double excitation operator that excites twp electrons from occupied spin orbital $i, j$ to virtual spin orbital $a, b$\\
    $i,j,k,l, \cdots$ & Subscribes of occupied spin orbital\\
    $a,b,c,d, \cdots$ & Subscribes of unoccupied (virtual) spin orbital\\
    $p,q,r,s, \cdots$ & Subscribes of spin orbital

    \end{tabular}
    
  \end{table*}

\end{document}